\documentclass[12pt]{article}

\usepackage{amssymb,amsmath}
\usepackage{openbib}
\usepackage{amsthm}
\usepackage[dvips]{graphicx}
\theoremstyle{plain}
\newtheorem{lem}{Lemma}
\newtheorem{prop}{Proposition}
\newtheorem{cor}{Corollary}
\newtheorem{theor}{Theorem}

\newtheorem{ex}{Example}

\def\ola{\overleftarrow}
\def\ora{\overrightarrow}
\def\cf{{\it cf. }}

\def\ie{i.e.\ }

\def\R{{\mathbb{R}}}      
\def\id{{id}}             
\def\H{\mathcal{H}}       
\def\D{\mathcal{D}}       
\def\B{\mathcal{B}}       
\def\J{\mathcal{J}}       
\def\I{\mathcal{I}}       
\def\P{\mathcal{P}}       
\def\LL{\mathcal{L}}      
\def\RR{\mathcal{R}}      
\def\M{\mathcal{M}}       
\def\SF{\mathcal{F}}      
\def\SSF{\mathfrak{F}}    

\def\df{\stackrel{\mathrm{def}}{=}}
\def\Do{\stackrel{\circ}{\Delta}}

\def\tr{\mbox {\rm tr} }
\def\Tr{\mbox {\rm Tr} }

\def\q{\hat q}
\def\p{\hat p}
\def\f{\hat f}
\def\h{\hbar}

\def\pr{\partial}
\def\L2{L^2(\R^n)}          

\def\d{{\delta^j_k}}
\def\g{\hat g}

\def\F{{F_{kj}}}
\def\h{\hbar}

\def\TB{T^*\!\M}        
\def\xb{\underline{x}}
\def\yb{\underline{y}}
\def\zb{\underline{z}}
\def\N{(2\pi \h)^n}     
\def\NN{\mathcal{N}}    
\def\MM{\M\times_{\xb}\M}
\def\theequation{\arabic{section}.\arabic{equation}}

\begin{document}
\begin{titlepage}

\font\title=cmbx12 \centerline{{}} \vspace{1cm}
 \centerline{{\title
Cotangent bundle quantization: Entangling of metric and magnetic field }}
  \vspace{10 mm}
 \centerline{M. V. Karasev } \vskip 8pt
\centerline{{\sl Department of Applied Mathematics}} 
\centerline{{\sl Moscow Institute of Electronics and Mathematics}} 
\centerline{{\sl Moscow 109028, Russia}}
\centerline{E-mail: karasev@miem.edu.ru}

\vspace{10mm} \centerline{T. A. Osborn } \vskip 8pt \centerline{{\sl Department
of Physics and Astronomy}} \centerline{{\sl University of Manitoba}}
\centerline{{\sl Winnipeg, MB, Canada, R3T 2N2 }}
\centerline{E-mail: tosborn@cc.umanitoba.ca}

\vspace{10 mm}

\begin{abstract}
\vspace{3 mm}

For manifolds $\M$ of noncompact type endowed with an affine connection (for
example, the Levi-Civita connection) and a closed 2-form (magnetic field) we
define a Hilbert algebra structure in the space $L^2(\TB)$ and construct an
irreducible representation of this algebra in $L^2(\M)$. This algebra is
automatically extended to polynomial in momenta functions and distributions.
Under some natural conditions this algebra is unique. The non-commutative
product over $\TB$ is given by an explicit integral formula. This product is
exact (not formal) and is expressed in invariant geometrical terms. Our
analysis reveals this product has a front, which is described in terms of
geodesic triangles in $\M$. The quantization of $\delta$-functions induces a
family of symplectic reflections in $\TB$ and generates a magneto-geodesic
connection $\Gamma$ on $T^*\M$. This symplectic connection entangles, on the
phase space level, the original affine structure on $\M$ and the magnetic
field. In the classical approximation, the $\h^2$-part of the  quantum
product contains the Ricci curvature of $\Gamma$ and a magneto-geodesic
coupling tensor.

  \vfill

\end{abstract}
\end{titlepage}

\section{Introduction}

There is a well-known quantum product defined by Groenewold
\cite{Gro46}, Moyal \cite{Moy49}, and Berezin \cite{FAB72} over
the phase space $\R^{2n}$. This non-commutative associative
product of functions corresponds to the Weyl symmetrization rule
for ordering the quantum coordinates $\q^j$ and momenta $\p_k$
which obey the canonical commutation relations
\begin{equation}\label {I.1}
[\q^j,\q^k]= 0\,,  \qquad [\q^j,\p_k]=i\h\delta^j_k \,, \qquad[\p_j,\p_k]=0\,.
\end{equation}
In the classical limit $\h \rightarrow 0$ this quantum product reduces to the
usual product of functions over $\R^{2n}$ and yields the standard symplectic
structure $dp \wedge dq $ on $\R^{2n}$.

There is also a magnetic analog \cite{KO1, KO3} of this quantum
product, where $\p$ plays the role of the kinetic momenta and
satisfies the commutation relations
\begin{equation}\label {I.2}
[\q^j,\q^k]=0\,, \qquad [\q^j,\p_k]=i\h\d\,, \qquad [\p_j, \p_k]=i\h\F(\q)\,.
\end{equation}
Here $F$ is the Faraday tensor (the strength of the magnetic field
for $n=3$).  The small
$\h$ asymptotics of this product generates the `magnetic'
symplectic structure on $\R^{2n}$
\begin{equation}\label{I.3}
\omega = d p\wedge dq + \frac 12 F(q)\, dq\wedge dq\,.
\end{equation}

The magnetic algebra generated by relations (\ref{I.2}) is  an
interesting and useful object for physical and mathematical
applications, [6--15].
In particular, the case of quadratic magnetic field $F$ represents
an example of quadratic quantum algebra (\ref{I.2}) which
corresponds to the symplectic space $\R^{2n}$ of constant non-zero
curvature \cite{KO3}. Also it is useful to recall that the form of
relations (\ref{I.2}) is gauge invariant, \ie it does not
depend on the choice of magnetic potential, and so all the
calculations in this algebra are gauge-independent a priori.

 One would like to examine what happens in this framework if the
flat $q$-space $\R^{n}$ is replaced by a curved manifold $\M$.
This means that the Euclidean metric on $\R^n$ is replaced by a
Riemannian metric, or more generally, by an affine
connection \underline{$\Gamma$} on $\M$. Accordingly, the phase
space $\R^{2n}$ is replaced by the cotangent bundle $\TB$.

The fundamental question arises: how to define a quantum product
over $\TB$ which  would naturally generalize the products
appearing in the Euclidean cases (\ref{I.1}) and (\ref{I.2}) and
incorporate the connection $\underline{\Gamma}$ on $\M$?

This is an old quantization problem, which was posed by Dirac
\cite{DR64} and Mackey \cite{Mac63} and initially studied in
[18--24]
and other works.

The recent mathematical investigation of this problem has been
carried out in [25--33]
including the case where the metric
and the magnetic field are both present
on $\M$ \cite{Liu92,Lan93,BPW03}.
There is
also a large literature, beginning with the paper by Widom
\cite{Wid80}, where this problem was studied from the perspective
of pseudodifferential and  Fourier integral operators.

In spite of certain essential progress, there remain many
significant open questions in this problem area.

First, we note that the papers cited above
do not address the following questions:

{\emph{How does the connection}} $\underline{\Gamma}$ {\emph{entangle with the
magnetic field}} $F$ on $\M$ via the quantization process? Are
$\underline{\Gamma}$ and $F$ combined in a natural geometrical way?

This family of questions closely parallels the issues raised in Weyl's
discovery of the gauge principle. In the paper \cite{Weyl18} Weyl constructed a
connection on the configuration space which combined the Levi-Civita metric
connection with the magnetic potential and was `gauge' invariant. Subjected to
Einstein's criticism \cite{Str96} that the construction was incompatible with
physical reality,  Weyl revised the direction of his program (and soon invented
the beginnings of modern gauge theory). Nevertheless we now think that Weyl's
original intention finds very strong support in the quantization theory where
one has a natural opportunity to extend configuration space $\M$ to the phase
space $\TB$ and re-examine the `connection problem' therein.

Secondly, although all the works dealing with the quantization problem over
$\TB$ use more or less the same core idea for generalizing the
Groenewold--Moyal product (just replace in the phase functions all the straight
chords by geodesics) there is a wide variation in the definition of the
amplitude functions. This variety of amplitudes illustrates the known phenomena
of the non-uniqueness of quantization.

Even in the Euclidean example an aspect of this non-uniqueness
is present, and correlates, for instance with the ordering
problem, \cf \cite{AW70a,KN78,KO1}. 
In the Euclidean case, conditions which uniquely identify the
Groenewold--Moyal product and Weyl ordering are known [39--41];
in the context of formal deformation theory see the discussion
in [42--44].  
The main idea in all these approaches is to exploit certain
symmetry group actions. 
For inhomogeneous manifolds $\M$ this is not possible. So, the question remains
open: How to select a {\emph{unique quantization}} on $\M$?

Thirdly, one may claim that in the literature there is still no
explicit formula for the quantum product over $\TB$, even for the
simplest examples of curved manifolds $\M$, even with no magnetic
field.

We mean here an {\emph{exact formula}}, not a formal deformation
one. Such an exact formula could be applied, for instance, to
highly oscillating or singularly concentrated (as $\h \rightarrow
0$) functions on $\TB$ which are required to describe
Schr\"odinger quantum dynamics or eigenfunction problems on $\M$.
On this topic we recall the asymptotic quantization theory
\cite{KM84} which allows this type of `semiclassical'
$\h$-dependence in its symbols and deals with symplectic manifolds
of general type without having a global polarization. However for
symplectic manifolds, $\prec\!\!\TB,\omega\!\!\succ$, it is
natural to ask more: namely to obtain an exact, not semiclassical,
quantization formula which is globally and geometrically stated on
$\TB$.

In this paper we present solutions to these questions in the case
where the configuration manifold  $\M$ is geodesically simply
connected. As an example one can take $\M$ to be a symmetric
Riemannian manifold of noncompact type, say, the hyperboloid in
the Minkowski space and, in particular, the Lobachevski  plane.
Another class of examples is given by manifolds $\M\approx \R^n$
whose metric is a deformation of the Euclidean one.

Part of the results described below can also be applied to generic
curved manifolds $\M$, for instance to compact manifolds.

The magnetic field $F$ can be an arbitrary closed 2-form on $\M$.

By using the averaging of $F$ along geodesics we define symplectic
transformations of the phase space $\prec\!\!\TB,\omega\!\!\succ$ which
correspond to autoparallel vector fields on $\M$. This is a
{\emph{magneto-geodesic analog of the Gallilei translations in $\R^n$}}. We
introduce unitary operators in $L^2(\M)$ corresponding to these symplectic
transformations and exploit them to select in a unique way the quantization
operation
\begin{equation}\label{I.4}
    f \rightarrow \f
\end{equation} on a function space over $\TB$ (Sect. 3).

The mapping (\ref{I.4}) determines an exact irreducible representation of the
Hilbert algebra $L^2(\TB)$ in the Hilbert space $L^2(\M)$. We also extend this
mapping to a wider algebra which includes, in particular, functions on $\TB$
polynomial in momenta, some exponential highly oscillating functions as
$\h\rightarrow0$, delta functions, etc. Note that we are employing here the
Hilbert algebra approach to quantization theory.  If one examines the
$C^*$-algebra corresponding to our quantum Hilbert algebra over $\TB$ then it
is of the strict quantization type \cite{Rief93}.

At the next stage in Sect.~4, we analyze which symplectic transformations
$\sigma_x$ of the phase space $\prec\!\!\TB,\omega\!\!\succ$ correspond to the
quantum $\delta$-functions, $\widehat \delta_x$.  In this way  a family of
{\emph{magneto-geodesic reflections}} on the space $\TB$ is obtained. They are
the phase space analogs of the geodesic reflections in $\M$ interacting with
the magnetic field.

In Sect.~5 we use the related $\sigma$-reflective curves to
represent the quantum product $\star$ over $\TB$ which corresponds
to the quantization operation (\ref{I.4}) in the usual way
\begin{equation} \label{I.7}
\f\, \g = \widehat{f {\star}\, g}\,.
\end{equation}
The product $\star$ is given by an exact, explicit and
geometrically invariant integral formula (\ref{I.5}).

This formula can be used for different subalgebras of functions over
$T^*\M$. Being restricted to the subalgera of polynomial in momenta
functions, this formula works for the case of a generic affine
manifold $\M$ (possibly not geodesically simply connected).

In Sect.~6 we prove that the asymptotic expansion of the exact
quantum product
as $\h \rightarrow 0$ has the following form
\begin{equation} \label{I.6}
f\,{\star}\,g =fg - \frac {i\h}2 f\langle
\ola\nabla\Psi\ora\nabla\rangle g - \frac {\h^2}{8} f\Big[ \langle
\ola\nabla\Psi\ora\nabla\rangle^2  + 3\langle
\ola\nabla\Psi\RR\Psi\ora\nabla\rangle\Big ]g+ O(\h^3)\,.
\end{equation}
Here $ \Psi =\big[\begin{array}{cc}0&-I\\I&F\end{array}\big]$ is the Poisson
tensor on $\TB$ associated with the symplectic structure (\ref{I.3}), $\nabla$
denotes the covariant derivative corresponding to a symplectic connection
$\Gamma$ on $\prec\!\!\TB,\omega\!\!\succ$ defined by magneto-geodesic
reflections, \cf (\ref{VI.5}) and  $\RR$ is
the Ricci tensor of this connection. The phase space
covariant derivative $\nabla$ appearing in the asymptotic expansion (\ref{I.6})
matches our quantization formulas with the deformation quantization
[18, 22, 28, 29, 42, 47--50]. 
In the deformation quantization framework the symplectic
connection corresponding to a given star product is determined
by the $\h^2$-term via a formula like (\ref{I.6}).

In our approach the connection $\Gamma$ is derived in a different
way, via $\sigma$-reflections.  In Sect.~6 the explicit formulas
are obtained for the connection $\Gamma$ and for its curvature in
terms of $\underline{\Gamma}$\, and $F$. These formulas entangle
the configuration space data $\underline{\Gamma}$\, and $F$ on the
phase space level. We call $\Gamma$ a {\emph{magneto-geodesic
connection}}.

The part of $\Gamma$ which depends on the magnetic field $F$ we
call a magneto-geodesic coupling. This is a 3-tensor on $\M$.
In the case of a Riemannian manifold $\M$ with a Levi-Civita
connection $\underline{\nabla}$ this tensor is equivalent to the
one which arises in the inhomogeneous Maxwell equation (with
current and charge).

The preprint version of this paper is found in arXiv:
quant-ph/0505144.

\section{Preliminary Definitions and Notation}
\setcounter{equation}{0}\setcounter{lem}{0}

After von Neumann \cite{vN27}, Wigner \cite{Wig32}, Groenewold
\cite{Gro46}, and Stratonovich \cite{STR57} it was understood that
the basic object of the quantization theory is a Hilbert algebra
together with its exact irreducible representation in a Hilbert
space.

By definition (see, for instance, in  \cite{Dix69}) the Hilbert algebra $\LL$
is a complete linear space with three structures: an associative product
$\star\,$, a scalar product $(\cdot,\cdot)$, and an involution $^*$, which are
mutually consistent.

Let $\H$ be a Hilbert space. Then the minimal Hilbert algebra
which has an irreducible representation in $\H$ is the algebra of
all Hilbert--Schmidt operators on $\H$.

The basic idea of quantization theory is to replace the operator
algebra by a function algebra over an appropriate phase space.
Following the Correspondence Principle one is taking $\H$ to be
\begin{equation*} \H =
L^2(\M,dm)\,,\end{equation*} where $\M$ is a configuration space,
\ie a smooth manifold with a smooth positive measure $dm$. The
phase space is then defined as $\TB$, \ie the cotangent bundle
over $\M$, and the Hilbert algebra is assumed to be
\begin{equation*}
\LL = L^2(\TB, dl)\,.
\end{equation*}
Here $dl$ is the normalized Liouville measure
\begin{equation*}
dl(x) = \frac {dx}{(2\pi \h)^{n}}\,, \qquad dx= dq^1\cdots dq^n dp_1\cdots
dp_n\,,
\end{equation*}
where $x=(q,p)\,,\, q\in \M\,,\, p\in T^{\star}_q\M\,,\, n={\rm\,
{dim}}\, \M$. So, the scalar product in the algebra $\LL$ is given
by
\begin{equation*}
(f,g) = \frac 1\N \int_{\TB} f(x)\, \overline{g(x)}\,dx\,,
\end{equation*}
and the involution is given by the complex conjugation
\begin{equation*}
f^* = \overline{f}\,.
\end{equation*}

In addition to the scalar product there is the trace functional
\begin{equation*}
\tr\,(f) \equiv \frac 1\N \int_{\TB} f(x)\, dx
\end{equation*}
where $f\in L^1(\TB, dl)$. We ask that the product in the algebra
$\LL$ obey the following property: the ideal $\LL^1 \equiv \LL
{\star}\, \LL$ is a subset of $L^1(\TB,dl)$, and
\begin{equation} \label{II.1}
\tr\,(f{\star}\,\overline{g}) = (f,g)
\end{equation}
for any $f,g\in \LL$.

The representation of the algebra $\LL$ in the Hilbert space $\H$
is denoted by
\begin{equation} \label{II.2}
f \rightarrow \f
\end{equation}
and is assumed to satisfy the usual axioms
\begin{equation}\label{II.3}
\f^\dag = \widehat{ \overline{f}}\,, \qquad \f \,\g =
\widehat{f{\star}\,g}\,, \qquad \Tr\,(\f) = \tr\,(f)\,.
\end{equation} Here $\Tr$ indicates the operator trace and $\dag$ the adjoint.
The last axiom is restricted to the subspace $\LL^1$.

The inverse to the mapping (\ref{II.2}) is called dequantization
or symbol mapping
\begin{equation*}
    \f \rightarrow f = {\mathrm {Smb}}(\f)\,.
\end{equation*}

It is convenient to write the quantization mapping (\ref{II.2}) in
the integral form
\begin{equation}\label{II.4}
\f\, = \int_{\TB }  f(x)\, \Delta_x\, dx\,,
\end{equation}
where $\{\Delta_x\}$ is a family of operators in $\H$ parameterized
by points $x \in \TB$. Then the symbol mapping is given by
\begin{equation}\label{II.5}
f(x)= (2\pi\h)^n \,\Tr\, (\f\, \Delta_x)\,.
\end{equation}

The first and third axioms in (\ref{II.3}), together with (\ref{II.1}), as well
as the definition (\ref{II.4}) together with (\ref{II.5}) are reformulated in
terms of the operator family $\Delta$ as follows
\begin{equation}\label{II.6}
\Delta_x^\dagger = \Delta_x\,, \quad (2\pi\h)^n\, \Tr\,(
\Delta_x\Delta_y ) = \delta_x(y)\,, \quad \N\int_{\TB}
\Delta_x\otimes\Delta_x \,dx = \I\,.
\end{equation}
Here $\delta$ is the Dirac delta-function with respect to the canonical measure
on $\TB$ and $\I$ is the antipodal operator $\psi\otimes\chi\rightarrow
\chi\otimes \psi$ on Hilbert space $\H\otimes \H$.

The second axiom in (\ref{II.3}) reads
\begin{equation}\label{II.7}
(f\,{\star}\,g)(x) = \int_{\TB}\int_{\TB} K_\star(x,y,z)\, f(y)\, g(z)\,dy \,d
z
\end{equation}
where the distribution $K_\star$ is defined as
\begin{equation}\label{II.7a}
K_\star(x,y,z) = (2\pi\h)^n\, \Tr\,(\Delta_x \Delta_y \Delta_z)\,.
\end{equation}

Of course, in formulas (\ref{II.4})--(\ref{II.7a}) appropriate care must be
taken with respect to the convergence of integrals and traces (using the weak
topology and a suitable distribution extension of functions). For instance, the
distribution character of $\Tr\, (\Delta_x \Delta_y) $ and $\Tr\, (\Delta_x
\Delta_y \Delta_z) $ is defined by first integrating the operator-valued
functions with $C_0^\infty$ test functions and after that computing the trace.

Various symmetry properties follow directly from the definition of $K_\star$ in
(\ref{II.7a}): it is invariant under any cyclic permutation of its arguments,
and it obeys $K_\star \rightarrow\overline{K_\star}$ under the permutation of
any pair of its arguments.

The family $\Delta$ was introduced by Stratonovich \cite{STR57} for the case
$\M=\R^n$. In \cite{VGB89} such a family was called a {\emph{quantizer}}. 
See also details and examples in [56--60].

One can call the last two quantizer properties in (\ref{II.6}) {\emph
{orthonormality}} and {\emph {operator completeness}}, respectively.  The
identities in (\ref{II.6}) imply that quantization (\ref{II.4}) and
dequantization (\ref{II.5}) are mutually consistent.

In the next section we construct the quantizer using the affine connection and
the magnetic field on $\M$, and then apply formulas (\ref{II.7}), (\ref{II.7a})
to calculate the quantum product over $\TB$.

But before that, we need to demonstrate how the ${\star}$ product can be extended to
other classes of symbols beyond the algebra $\LL$.

Let $\SF$ be any subalgebra in $\LL$. Denote by $\SF'$ the space of linear
functionals on $\SF$.  Employing the canonical measure we identify functionals
with distributions on $\TB$ via
\begin{equation*}
\langle f,h \rangle = \int_{\TB} f(x)\,h(x)\,dx\,, \qquad f\in \SF'\,, h\in
\SF\,.
\end{equation*}
Obviously $\SF\subset \SF'$. Further note that $\SF'$ is an $\SF$-module, \ie $
f\in\SF'\,,\,k\in \SF \quad \Rightarrow \quad f\,{\star}\,k \in \SF'\,,\,\,
k\,{\star}\,f\in \SF'$ where by definition
\begin{equation}\label{II.7b}
\langle f\,{\star}\,k, h\rangle \df \langle f,k\,{\star}\, h\rangle\,, \qquad
\langle k\,{\star}\,f, h\rangle \df \langle f,h\,{\star}\, k\rangle\,, \qquad
\forall\, h \in \SF\,.
\end{equation}
Denote by $\SF_{\star}$ the following subset
\begin{equation*}
\SF_{\star} = \{f \in \SF'|\, f{\star}\,h \in \SF\,\, {\mathrm{and}}\,\,
h\,{\star}\,f \in \SF\,,\,\, \forall\, h\in \SF \}\,.
\end{equation*}
In particular, $\SF\subset \SF_{\star}$\,.

We call $\SF$ a {\emph{normal subalgebra}} if the set $\SF_{\star}$ obeys the
property
\begin{equation} \label{II.9}
\langle f, h\,{\star}\,g\rangle = \langle g, f\,{\star}\,h \rangle\,, \qquad
\forall\, f,g \in \SF_{\star}\,, \quad \forall\, h\in \SF\,.
\end{equation}

If $\SF$ is a normal subalgebra in $\LL=L^2(\TB)$, then the set
$\SF_{\star}$ is endowed with the algebra structure
\begin{equation} \label{II.10}
\langle f{\star}\,g, h\rangle \df \langle f, g\,{\star}\,h \rangle\,, \qquad
\forall\, f,g \in \SF_{\star}\,, \quad \forall\, h\in \SF\,,
\end{equation}
which is consistent with the involution $\overline{f\,{\star}\,g}=
\overline{g}\,{\star}\,\overline{f}\,.$

Verification of the embedding $f\star g \in \SF_{\star}$ and the
$\star$-associativity is achieved by repeated applications of
(\ref{II.7b})--(\ref{II.10}) in combination with the associativity
of $\SF$. The algebra $\SF_{\star}$ is a natural extension of the
subalgebra $\SF$.

Note that the unity function $1$ does not belong to $\SF$ or $\LL$, but is
automatically an element of $\SF_{\star}$ and $
1\,{\star}\,f = f\,{\star}\,1 = f\,, \  \forall\, f \in \SF_{\star}\,.
$
 So, $\SF_{\star}$ is an involutive algebra with unity.

In the next section we introduce a concrete example of a normal
subalgebra $\SF \subset \LL$ and its extension $\SF_*$ suitable
for quantizing the phase space $\TB$. In the case $\M=\R^n$
similar extensions were used in [15, 61--64].

\section{Quantization and dequantization over $\TB$}
\setcounter{equation}{0}

Let $\M$ be a smooth oriented manifold with an affine torsion free
connection $\underline{\Gamma}$ and a smooth positive measure $dm$.

We assume that $\M$ is geodesically simply connected, that is, every pair of
points is connected by a unique geodesic, and moreover this geodesic is
infinitesimally isolated (has no conjugate points).

For any $q\in\M$ we use the notations
\begin{eqnarray*}\label{III.1}
V_q &=& {\underline{\exp}}_{\ q}^{-1}\,, \qquad  s_q = {\underline{\exp}}_{\ q}(-V_q)\,,\hfill\\
j_q &=& 2^{-n}|\det\big(\pr V_q + \pr V_q(s_q)\big)|\,,\qquad \pr
V_q \equiv \frac {\pr V_q}{\pr q}\,,  \phantom{\bigg (}
\\
J_q &=& j_q\, \frac {\D m(s_q)}{\D m}\,,  \qquad e_q \equiv \frac {\D
m({\underline{\exp}}_{\ q}(v))}{\mu(q) \D v}\,\bigg|_{v=V_q}\,.
\end{eqnarray*}
In these formulas one has the following objects:
\begin{itemize}
    \item the exponential map ${\underline{\exp}}_{\ q}:T_q\M \rightarrow \M$  which is
    everywhere non-degenerate,
    \item for any $q'\in \M$ the vector $V_q(q') \in T_q\M$ is the velocity
    on the geodesic connecting $q$ with $q'$ in unit time,
    \item the mapping $s_q:\M \rightarrow \M$ is the geodesic reflection about point $q$, ${s_q}^2 =\id$\,,
    \item the Jacobian $j_q\in C^\infty(\M)$ is invariant under
    the reflection $s_q$\,,
    \item $\mu>0$ denotes the density  of the measure on $\M$, so that  $dm(q) =
    \mu(q)\,dq$\,,
    \item the Jacobian $\D m(s_q)/\D m \in C^\infty(\M)$ is obtained
    by transforming the measure $dm$ under the diffeomorphism $s_q$\,,
    \item the Jacobian $e_q \in C^\infty(\M)$ determines the
    transformation of the measure $dm$ under the diffeomorphism ${\underline{\exp}}_{\ q}$\,.
\end{itemize}

In addition, let $F$ be a closed 2-form on $\M$. We fix an
arbitrary point $o\in\M$ and define the function $\Phi_q \in
C^\infty(\M)$ as
\begin{equation} \label{III.2}
\Phi_q(q') = \int_{\pi_q(q')} F\,.
\end{equation}
Here $\pi_q(q')$ is a two-dimensional surface in $\M$ whose
oriented boundary is composed of three geodesics (Fig. 1): the
geodesic from $q'$ to $o$, from $o$ to $s_q(q')$, and from
$s_q(q')$ back to $q'$. The values of the function $\Phi_q$ are
just the magnetic flux through the surfaces $\pi_q$.
\begin{figure}\label{figA}
\centering
\includegraphics{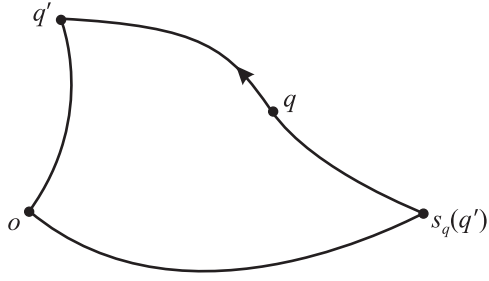}
\caption{Geodesic triangle $\pi_q(q')$ in $\M$.}
\end{figure}

\bigskip Now we are ready to define
the quantizer $\{\Delta_x|\, x\in \TB\}$. This family of operators
is determined by its integral kernels $\Delta_x(a,b)$
\begin{equation*}
(\Delta_x\psi)(a) = \int_{\M} \Delta_x(a,b)\,\psi(b)\,dm(b)\,,
\qquad \psi\in \D(\M)\,.
\end{equation*}
Here and below we denote by $\D(\NN)\equiv C_0^\infty(\NN)$ the
space of all compactly supported $C^\infty$-functions on a
manifold $\NN$.

Let $x=(q,p)$, so that $p\in T^*_q\M$. We set
\begin{equation} \label{III.3}
\Delta_x(a,b) \df \frac 1{(\pi\h)^n} \sqrt{J_q(a)} \exp\Big\{ \frac
i\h\big(2p\,V_q(a) + \Phi_q(a)\big)\Big\}\,\delta_{s_q(a)}(b)\,,
\end{equation}
where $\delta$ is the delta-function on $\M$ with respect to the measure $dm$;
the product $p\, V_q(a)$ represents the natural pairing of the covector $p$ and
the tangent vector $V_q(a)$ in $T_q\M$.

\begin{lem} \label{lem1} The family of operators $\{\Delta_x\}$
defined  by the integral kernels
{\rm(\ref{III.3})} obeys properties {\rm(\ref{II.6})} and so is a
quantizer{\,\rm :}
\begin{equation} \label{III.6}
\begin{array}{c} {\displaystyle{
\overline{\Delta_x(a,b)} = \Delta_x(b,a)\,,}} \\
{\displaystyle{ \N\int_{\M}\int_{\M}\Delta_x(a,b)\,\Delta_y(b,a)\,dm(a)\,dm(b)
= \delta_x(y)\,,}} \phantom{\Bigg (}\\ {\displaystyle{
\N\int_{\TB}\Delta_x(a,b)\,\Delta_x(c,d)\, dx = \delta_a(d)\,\delta_b(c)\,. }}
\end{array}
\end{equation}
\end{lem}

The proof follows directly from the definition (\ref{III.3}) by
simple computation of integrals containing delta-functions.

For some manifolds $\M$ the function $j_q$ may be unbounded. In this case the
quantizer $\Delta_x$ is an unbounded operator but remains selfadjoint with
domain $\D(\Delta_x) = \{\psi \in \H\,|\int j_q |\psi|^2\,dm\ < \infty\}$. The
family $\{\Delta_x\}$ and all $x$-derivatives are strongly continuous on
$\D(\M)$.

   Recall that the operators $\hat f$ are defined by formula (\ref{II.4}).
Let $\SSF$ denote the integral kernel of the operator $\f$,
\begin{equation} \label{III.8a}
(\f\psi)(a) = \int_{\M} \SSF(a,b)\,\psi(b)\,dm(b)\,, \qquad \psi \in \D(\M)\,.
\end{equation}\smallskip
For simplicity we assume that $\SSF\in \D(\M \times \M)$.

Applying formula (\ref{II.5}), one obtains the analog of the Wigner transform.
\begin{lem}The symbol $f$ is constructed from the kernel
$\SSF$ via
\begin{equation} \label{III.9}
f(q,p) = \int_{T_q\M} e^{\textstyle{-i up/ \h}}\,
\big(k\SSF\big)\big({\underline{\exp}}_{\ q}( {\frac u2} )\,,
{\underline{\exp}}_{\ q}(-\frac u2) \big)\,\mu(q)\,du\,.
\end{equation}
\noindent Here the function $k$ is given by
\begin{equation} \label{III.10}
k(a,b) \df \Big(j_{a \vee b}(a)\,e_{a \vee b}(a)\,e_{a \vee b}(b)\Big)^{1/2}\,
\exp\Big\{\textstyle{-i \Phi_{a \vee b}(a)/ \h}\Big\}\,,
\end{equation}
\noindent and $a \vee b$ denotes the geodesic midpoint {\rm(}Fig.~{\rm2)},
that is
\begin{equation} \label{III.10a}
s_{a\vee b}(a) = b\,.
\end{equation}
\end{lem}

The inverse transform from symbol to kernel results from (\ref{II.4}). Denote
the Fourier image of $f$ in the momentum variable by
\begin{equation} \label{III.10b}
f^\sim(q,u) \df \frac1\N \int_{T_q^{\star} \M} e^{\textstyle{i up/
\h}}\,f(q,p)\, \frac {dp}{\mu(q)}\,.
\end{equation}

\begin{lem} \label{3.3} The integral kernel $\SSF$ of the operator corresponding to the
symbol $f\in \D(T^*\M)$, is given by
\begin{equation}\label{III.11}
\SSF(a,b) = \frac {f^\sim(a\vee b, a\wedge b)}{k(a,b)}\,,
\end{equation} where $a \vee b $ is the geodesic midpoint
{\rm(\ref{III.10a})} and $a \wedge b$ is
the geodesic velocity at the midpoint {\rm(}Fig.~{\rm2)}
\begin{equation}\label{III.12}
a\wedge b \df V_{a\vee b}(a) - V_{a\vee b}(b) = 2\,V_{a\vee b}(a)\,.
\end{equation}
\end{lem}
In the flat case  $\M=\R^n$ one has $|k(a,b)|=1$ and (\ref{III.9}),
(\ref{III.11}) becomes the standard Wigner transform in the presence
of a magnetic field.

Using (\ref{III.11}) one can readily compose the product of two
operators and find a simple composition rule in terms of
Fourier-imaged symbols. To formulate the result let us recall that
the tangent bundle $T\M$ is endowed with a natural groupoid
multiplication \cite{Con94}
\begin{equation}\label{III.12a}
    n',n'' \mapsto n'\circ n''
\end{equation}
by means of the left and right (target and source) mappings
\begin{eqnarray*}\label{III.12b}
\tilde l:T\M &\rightarrow& \M\,, \quad \tilde l(n) \df {\underline{\exp}}_{\ q}(\frac u2)\,, \\
\tilde r:T\M &\rightarrow& \M\,, \quad \tilde r(n) \df
{\underline{\exp}}_{\ q}(-\frac u2)\,, \quad n\equiv (q,u) \in
T\M\,.
\end{eqnarray*}
Namely, the product (\ref{III.12a}) of two elements $n',n'' \in T\M$ is well
determined iff $\tilde r(n') = \tilde l(n'')$, and in this case one has $\tilde
l(n'\circ n'')=\tilde l(n')\,, \, \tilde r(n'\circ n'')=\tilde r(n'')$. Note
that the mappings $\tilde l, \tilde r$ themselves can be expressed in terms of
the groupoid multiplication as
\begin{equation*}
\tilde l(n) = n \circ n^{-1}\,, \qquad \tilde r(n) = n^{-1} \circ n\,,
\end{equation*} where $n^{-1}\equiv(q,-u)$ is the element inverse to $n$ in
$T\M$.

\begin{figure}
\centering
\includegraphics{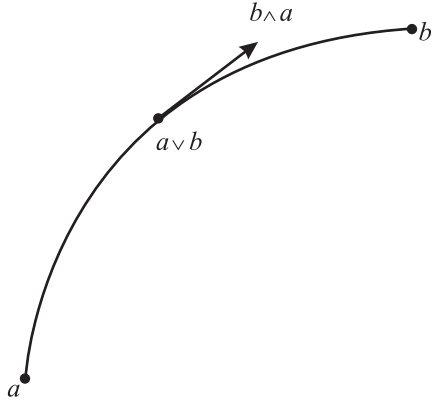}
\caption{Geodesic mid-point and velocity in $\M$.}
\end{figure}

\begin{lem} \label{3.4} Composition of Fourier--imaged symbols over $T\M$ is
given by the groupoid modified convolution
\begin{equation} \label{III.13}
\big(f^\sim \odot g^\sim\big)(n) = \int_{n=n'\circ n''}\frac
{\kappa(n)}{\kappa(n')\kappa(n'')}\,f^\sim(n')\,g^\sim(n'')\,
dm\,.
\end{equation}
Here $n,n',n''$ are points from $T\M$, the function $\kappa(n)\equiv k(\tilde
l(n), \tilde r(n))$ is given by {\rm{(\ref{III.10})}} and by the left and right
mappings of the groupoid structure {\rm(\ref{III.12a})}. The integration in
{\rm{(\ref{III.13})}} is taken with respect to the measure $dm \big(\tilde
r(n')\big) = dm \big(\tilde l(n'')\big)$ over the manifold $\M$.
\end{lem}

Formula (\ref{III.13}) belongs to the class of Connes' type tangential groupoid
quantization formulas \cite{Con94,Car99,KO1}. Note that in the convolution
integrand (\ref{III.13}) we have an additional {\it{groupoid cocycle}}
\begin{equation} \label{III.15}
C(n', n'') = \frac {\kappa(n'\circ n'')}{\kappa(n')\kappa(n'')} =
|C(n',n'')|\,\exp\left \{ \frac i\h  \int_{\Delta(n',n'')} F
\right\}\,.
\end{equation}
The cocycle property
\begin{equation}\label{III.15b}
C(n,m\circ l)\, C(m,l) = C(n\circ m, l)\, C(n,m)
\end{equation}
guaranties the associativity of the modified groupoid convolution
(\ref{III.13}).

The phase of the cocycle (\ref{III.15}) is just the magnetic flux
through the triangle $\Delta(n',n'')$ in $\M$ bounded by geodesics
(Fig.~3) with mid-points $q,q',q''$ and mid-velocities $u,u',u''$
such that
\begin{equation*}
n'=(q',u'), \qquad n''=(q'',u''), \qquad n'\circ n'' = (q,u)\,.
\end{equation*}
This phase is similar to its form in the Euclidean case \cite{KO1}, but now it
also senses the non-Euclidean connection on $\M$.  On the phase space level the
property (\ref{III.15b}) is equivalent to the Stokes theorem applied to the
geodesic tetrahedron in $\M$ with sides corresponding to elements $n,m,l,
m\circ l, n \circ m, l\circ n, n\circ m\circ l \in T\M$.

\begin{figure}
\centering
\includegraphics{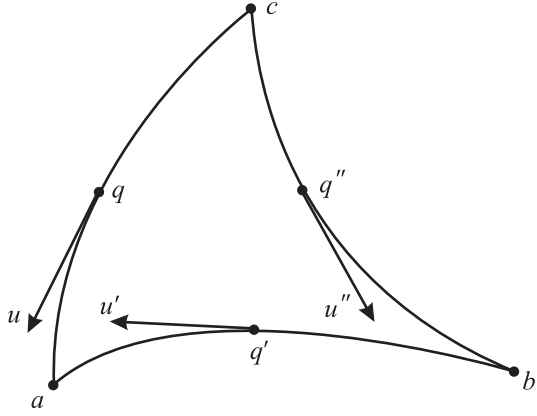}
\caption{Geodesic triangle in $\Delta(n',n'')$ in $\M$.}
\end{figure}

The amplitude $C(n', n'')$ of the cocycle (\ref{III.15}) obtained from
(\ref{III.10}) is
\begin{equation*}
|C(n', n'')| = \left( {\frac{j_q(c)}{j_{q'}(a)\,j_{q''}(b)}} \cdot
\frac {e_q(a) e_q(c)}{e_{q'}(a)e_{q'}(b) e_{q''}(b)e_{q''}(c) }
\right)^{1/2}\,.
\end{equation*}
This amplitude is an additional `geodesic' contribution to the groupoid cocycle
structure (\ref{III.15}).

Obviously, one has the following
\begin{prop}\label{lem4a}
 The quantization mapping $f\rightarrow
\f$, defined by {\rm{(\ref{III.8a})}}, {\rm{(\ref{III.11})}} is an isomorphism
between the algebra $\LL = L^2(\TB,dl)$ and the algebra of all Hilbert--Schmidt
operators acting on the Hilbert space $\H=L^2(\M,dm)$. The quantum product
$\star$ in the algebra $\LL$ is generated by the groupoid modified convolution
{\rm{(\ref{III.13})}}, as follows
\begin{equation}\label{III.16}
\big( f{\star}\,g \big)^\sim= f^\sim \odot g^\sim\,.
\end{equation}
\end{prop}

\bigskip
Now we would like to extend the quantization mapping to a wider
algebra.

Examining formula (\ref{III.13}) one can easily see that the class $\D(T\M)$ is
invariant with respect to the $\odot$ convolution. Moreover, the class of
$C^\infty$-functions on $TM$ with compact support in tangential directions is a
$\D(T\M)$-module with respect to this convolution.  Thus one is led to the
following definitions.

Denote by $\SF = \SF(\TB)$ the space of functions on $\TB$ whose momentum
Fourier image belongs to $\D(T\M)$.

Further, denote by $\P=\P(\TB)$ the space of $C^\infty$ functions
on $\TB$  which are polynomial in momenta.  Let $\P^l \subset\P$
be the subspace which consists of polynomials of degree $l$ in
momenta.

From the above discussion and the definitions of Sect.~2 we have
\begin{prop} \label{prop1} $\SF$ is a normal subalgebra of the Hilbert algebra
$\LL=L^2(\TB)$.
\end{prop}

\begin{prop} \label{prop2} The space $\P$ is an involutive
subalgebra in $\SF_\star$ with the gradation:\\
$\P^l {\star}\, \P^m \subset \P^{l+m}$.  The commutator in the algebra $\P$ is
graded as follows
\begin{equation}\label{III.17}
[\P^l,\P^m]_{\star} \subset \P^{l+m-1}\,, \qquad l,m\geq 0\,,
\end{equation}
where $\P^{-1}\equiv 0$. In particular, $[\P^0,\P^0]_\star = 0$.
\end{prop}

For any $f\in \SF_{\star}$ the operator
\begin{equation}\label{III.18}
\f: \D(\M) \rightarrow \D'(\M)
\end{equation}
is defined by its bilinear form
\begin{equation}\label{III.19}
(\f\psi, \chi)_{\H} = \langle f, \rho_{\psi|\chi}\rangle \,.
\end{equation}
Here $\rho_{\psi|\chi}$ is the {\it{Wigner function}}
\begin{equation}\label{III.20}
\rho_{\psi|\chi}(x) \df (\Delta_x \psi, \chi)_{\H}\,.
\end{equation}
It is evident that $\rho_{\psi|\chi} \in \SF$ if $\psi,\chi \in
\D(\M)$\, and so the `matrix elements' (\ref{III.19}) are
well-defined. In the particular case where $f\in \SF$ this definition
of the operator $\f$ coincides with the above definition
(\ref{III.8a}), (\ref{III.11}).

As one would expect, the Wigner function obeys the usual probability
interpretation together with the associated bound condition
\begin{equation*}
\int_{T^*_q\M}\rho_{\psi|\psi}(q,p)\, dp = |\psi(q)|^2\,, \quad
|\rho_{\psi|\psi}(q,p)| \leq {\frac 1{(\pi\h)^n}}\int_\M (j_q)^{1/2} |\psi|^2
dm\,.
\end{equation*}

\begin{lem} \label{lem6}
If $f\in \P$ then $\f$ maps $\D(\M)$ into $\D(\M)$; moreover, it
is a differential operator.  If $f\in \P^l$ then $\f$ is a
differential operator of order $l$.
\end{lem}

Let us consider now some basic examples of functions from
$\SF_\star$ or $\P$ and the operators (\ref{III.18}) corresponding
to them.
\begin{ex}

Let $1$ be the unit function on $\TB$. Then $1\in \SF_\star$ and
$\hat 1 = I$ is the identity operator in $L^2(\M,dm)$.

Let $\delta_x$ be the delta-function on $\TB$ concentrated at the
point $x$. Then $\delta_x \in \SF_{\star}$. The corresponding
operator is the quantizer {\rm{(\ref{III.3})}},
\begin{equation*}
\widehat{\phantom{,}\delta_x}  = \Delta_x\,.
\end{equation*}
\end{ex}
\noindent In the Euclidean case see \cite{Gro76,Roy77}.

\begin{ex}
Let $f\in \P^1$. Then $f$ is represented by the sum $f(q,p)=\varphi(q)+
f_W(q,p)$ with $\varphi \in C^\infty(\M)$, $f_W(q,p)= p\,W(q)$, and where  $W$
is a vector field on $\M$. Obviously $\widehat \varphi = \varphi$ (the
multiplication operator) and
\begin{equation}\label{III.21} \widehat{f_W} = -i\h\,(W +
\frac 12\, {\mathrm{div}}\, W) - A^o\, W\,.
\end{equation}
\end{ex}
\noindent Here on the right hand side the field $W$ is considered as a first
order differential operator, and div\,$W$ denotes the divergence of $W$ with
respect to the measure $dm$. The function $A^o\,W$ in (\ref{III.21}) is the
pairing of the vector field $W$ with the 1-form $A^o$ on $\M$, and $A^o$ is a
primitive of the Faraday 2-form, \ie
\begin{equation*}
dA^o = F\,.
\end{equation*}
This 1-form is uniquely determined by the radial gauge condition
that $A^o(q)$ is {\it{perpendicular to the velocity of the geodesic
connecting $o$ with $q$}};
\begin{equation} \label{III.21a} A^o(q)\, V_q(o) = 0\,.
\end{equation}
An explicit formula for $A^o$ is the following \cite{MOF90}
\begin{equation} \label{III.22}
A^o(q) = \int_o^q\left({\frac {\pr Q}{\pr q}}\right)^* F(Q)\,
dQ\,,
\end{equation}
where $Q$ is the geodesic from $o$ to $q$, and the integral
(\ref{III.22}) is taken along this geodesic. The magnetic potential
(\ref{III.22}) can be interpreted as the average of the Lorentz force
$F(Q)\, \dot Q$ with respect to the Jacobi field along the geodesic.

At this stage it is useful to discuss the relationship between
various gauge choices and structure of the quantizer. The definition
(\ref{III.3}) of $\Delta_x$ makes no reference to any particular
magnetic potential. However the magnetic phase flux $\Phi_q$, when
written as a line integral, is expressed in terms of the radial gauge
by
\begin{equation} \label{III.21b}
\Phi_q(a) = \int_{s_q(a)}^a A^o \,.
\end{equation}
Here the integral is taken along the geodesic on $\M$ connecting
$b\!=\!s_q(a)$ to $a$.  The $a\, o$ and $o\, b$ geodesic
contributions to $\Phi_q(a)$ vanish  by virtue of the radial gauge
condition (\ref{III.21a}). The Examples 2 and 3 demonstrate the
dependence on just this special radial gauge 1-form $A^o$.  If one
modifies the definition of the quantizer by replacing $A^o$ in
(\ref{III.21b}) by a magnetic vector potential in some other gauge,
say $A$, then the potential $A$ will replace $A^o$ in the
quantization of the momentum coordinate, (\ref{III.21}). We conclude
that although the quantum algebra is gauge invariant, its
representation depends on the gauge choice.

In the Euclidean case $\M=\R^n$ the 1-form (\ref{III.22}) coincides
with the Valatin potential \cite{Val54}, and condition
(\ref{III.21a}) is the Dirac gauge condition (see details in
\cite{KO1}).

\begin{ex}  Let $M$ be a covariantly constant bivector field on $\M$ and $f_M(q,p)\!
\equiv\! M^{jk}(q)p_jp_k$ $ \forall p \in T^*_q\M$. Then $f_M \in
\P^2$ and
\begin{equation*}
\widehat{f_M} = (-i\h \underline{\nabla}_j - A^o_j)\,M^{jk}\,(-i\h
\underline{\nabla}_k - A^o_k)\,,
\end{equation*}
where $\underline{\nabla}$ is the covariant derivative defined by
the connection $\underline{\Gamma}$ on $\M$, and $A^o_j$ are
components of the 1-form {\rm(\ref{III.22})}.
\end{ex}

In particular, let \underline{$\Gamma$} be the Levi-Civita
connection , and $g=(\!(g_{jk})\!)$ be the metric tensor on $\M$.
Then
\begin{equation*}
\widehat{g^{jk}(q)p_j p_k} = g^{jk}(-i\h \underline{\nabla}_j -
A^o_j)\,(-i\h \underline{\nabla}_k - A^o_k)\,.
\end{equation*}
If the magnetic field is absent this reduces to $-\h^2 \triangle$
(the Laplace operator on $\M$).

\begin{ex}  Let $W$ be a vector field on $\M$, and $r \in C^\infty(\M)$.
Consider the function on $\TB$ given by
\begin{equation}\label{III.23}
r_W(q,p) \equiv r(q)\exp\Big\{ {\frac i\h}p\,W(q) \Big\}\,, \qquad p \in
{T^*}_q\M\,.
\end{equation}
Suppose the following condition holds{\rm:}
\begin{equation} \label{III.24}
{\mathit{the\, matrix}}\,\, \frac 12 \pr W(b)/\pr b   {[\pr V_a(b)/\pr
b]}^{-1}\,\, {\mathit{has\, no\, eigenvalues\, \pm 1\, for\, any\, }}a,b \in
\M\,.
\end{equation}
Then $r_W \in \SF_{\star}$.
\end{ex}

Indeed, by computing ${r_W}^\sim$ and transforming this to the
kernel form (\ref{III.11}) one obtains the delta-function
$\delta\big(W(a\vee b) + a \wedge b\big)$, which must be a
well-defined distribution in both $a$ and $b$ (in order to achieve
the embedding $r_W \in \SF_\star$). Thus it is sufficient to
assume that both determinants
\begin{equation}
\det\, {\frac {\pr}{\pr a}}\big(W(a \vee b) + a \wedge b\big) \neq
0, \qquad \det\, {\frac {\pr}{\pr b}}\big(W(a \vee b) + a \wedge
b\big) \neq 0\,,
\end{equation} are non-zero for all $a,b \in \M$. This is equivalent to
condition (\ref{III.24}).

The operator  corresponding to the symbol $r_W$, subject to the condition
(\ref{III.24}), acts in $\H$ as
\begin{equation}\label{III.25}
(\widehat{r_W}\psi)(a) = {\frac {J_q\big({\underline{\exp}}_{\
q}(-\frac 12 W(q))\big)^{1/2}}{\Big| \det\big(\pr V_q(a) + \frac
12 \pr W(q)\big)\Big|} }\, r(q) e^{\textstyle{-i \Phi_{q}(a)/
\h}}\, \psi\big({\underline{\exp}}_{\ q}({\textstyle{\frac 12}}
W(q))\big)\bigg|_{q=Q(a)}
\end{equation}
where $Q(a)$ is the solution of the equation
\begin{equation}\label{III.26}
{\underline{\exp}}_{\ Q}\big({\textstyle{-\frac 12}} W(Q)\big) =
a\,.
\end{equation}

Now, consider the specific class of {\it autoparallel vector fields}, \ie those
satisfying the identity
\begin{equation*}
\underline{\nabla}_W W
 = 0 \qquad {\mathrm or} \qquad W^j\pr_jW^k
+ \underline{\Gamma}_{js}^k W^j W^s = 0\,.
\end{equation*}
In this case the flow $R^t :\M \rightarrow \M$ of the field $W$ is
given by
\begin{equation}\label{III.27}
R^t(q) = {\underline{\exp}}_{\ q}\big(t\,W(q)\big)\,.
\end{equation}
Note that condition (\ref{III.24}) is satisfied automatically and
equation (\ref{III.26}) now reads\\
$R^{-1/2}(Q) = a$, and so, $Q(a) = R^{1/2}(a)$.
\smallskip

Moreover, ${\underline{\exp}}_{\ q}\big(\frac 12 W(q)\big) =
\exp_a\big(W(a)\big)$ and so $\psi({\underline{\exp}}_{\
q}\big(\frac 12 W(q)\big)\big|_{q=Q(a)}\! = \psi(R^1(a))$.
\smallskip

\noindent Thus the operator $\widehat{r_W}$ acts as a shift
operator along the trajectories of the field $W$.
\begin{lem} \label{3.7} Let $W$ be an autoparallel vector field on $\M$,
generating the flow {\rm(\ref{III.27})}. Set
\begin{equation}\label{III.28}
r^t(q) \df \det\, DV_q(R^{t/2}(q)) \left( \frac {\det\, dR^{t/2}(q)
\cdot \det\, dR^{-t/2}(q)} {j_q\big(R^{t/2}(q)\big)}
\right)^{1/2}.
\end{equation}
Then the family
\begin{equation} \label{III.29}
\widehat{r_W^t} \equiv \widehat{ r^t \exp \big\{\frac {it}\h f_W
\big\} } \qquad (t\in \R)\,,
\end{equation}
where $f_W(q,p) \equiv p   W(q)$, forms a one-parameter group of unitary
operators in $\H$ acting by the formula
\begin{equation} \label{III.30} \big(\widehat{r_W^t}\psi\big)(a) =
\sqrt{\frac {\D m\big(R^t(a)\big)}{\D m(a)} }\exp\left \{ \frac
i\h \int_{\pi^t(a)} F \right\}\,\psi\big(R^t(a)\big)\,.
\end{equation} Here the boundary of a surface $\pi^t(a) \subset \M$ is composed
of three geodesics connecting the points $o\rightarrow R^t(a)
\rightarrow  a\rightarrow o$.
\end{lem}

Indeed the operator given by (\ref{III.30}) is automatically
unitary. So, to prove the lemma we just need to compare formulas
(\ref{III.30}) and (\ref{III.25}), and choose an appropriate
amplitude function $r$.

Note that the mapping $R^t: \M \rightarrow\M$ can be naturally
lifted up to the mapping
\begin{equation}\label{III.31}
\gamma_W^t : \TB \rightarrow \TB\,, \qquad
\gamma_W^t\left(\begin{array}{c} q \\ p\end{array}\right) =
\left(\begin{array}{l} R^t(q)\\{dR^t(q)}^{-1*}(p + \beta^t)
\end{array}\right)\,.
\end{equation}
Here the covector $\beta^t \in T^*_q \M$ is defined by $\beta^t =
A(q,R^t(q))$, where
\begin{equation}\label{III.32}
A(q,a) \df \int_a^q \left({\frac {\pr Q}{\pr q}}\right)^* F(Q)\,
dQ\,.
\end{equation}
The potential $A(q,a) \in T_q^*\M$ is the version of (\ref{III.22}) with the
initial point $a$ instead of $o$: $d_q\,A(q,a) = F(q),\,\, A(q,a)  V_q(a) = 0$.
In particular, if $a=o$, then $A(q,o)\equiv A^o(q)$ in the notation of
(\ref{III.22}).

\begin{cor}\label{cor1}  Let $W$ be an autoparallel vector field on $\M$, and
$g \in \P^1$. Then the following permutation formula holds on the dense domain
$\D(\M)\subset\H$
\begin{equation}\label{III.33}
\widehat{r^t_W}\cdot\widehat g \cdot {\widehat{r^t_W}}^{-1} =
\widehat{g^t_W}\,, \qquad g_W^t\df {\gamma_W^t}^*g\,.
\end{equation}
\end{cor}

Thus we see that the unitary operators $\widehat{r^t_W}$ with
symbols of exponential type (\ref{III.29}) play the role of the
quantum transformations corresponding to the classical symplectic
transformations $\gamma_W^t$ (\ref{III.31}). Permutation formula
(\ref{III.33}) belongs to the general class of Fock-type formulas
\cite{Fock} which relate classical symplectic transformations to
quantum unitary operators
(see also in \cite{KN78,KM84,Foll89}).

Properties (\ref{III.17}) and (\ref{III.33}) which we derived form
the definition of the quantizer (\ref{III.3}) actually determine
the quantization uniquely.

\begin{prop} \label{prop3} If a quantization obeys the axioms {\rm(\ref{II.6})} and the
graded commutator property {\rm(\ref{III.17})}, and if for any
autoparallel vector field $W$ on $\M$ there is a function $r^t \in
C^\infty(\M)$  such that the operator $\widehat{r_W^t}$
{\rm{(\ref{III.29})}} is unitary and the property
{\rm(\ref{III.33})} holds, then this quantization coincides with
the one defined by \rm{(\ref{III.9})}, \rm{(\ref{III.11})}.
\end{prop}

These conditions for uniqueness are, in a sense, analogous to
those known in the Euclidean case with no magnetic field
[39--41], but our Proposition {\ref{prop3} uses
different logical assumptions.

We call the quantization defined by ($\ref{III.9}$), $(\ref{III.11})$ a
{\it{magneto-geodesic quantization}}.

Observe that if the form of the quantizer kernel is a priori assumed to be of
the type (\ref{III.3}) having the phase and $\delta$-function structure given
there but with unknown amplitude, then requiring that $\Delta_x$ obey axioms
(\ref{II.6}) fixes the amplitude. The additional properties (\ref{III.17}),
(\ref{III.33}) in Proposition \ref{prop3} were introduced to uniquely select
the $\delta$-function and the phase function of (\ref{III.3}).

Note that in the case $F=0$ (no magnetic field) the quantization which we
uniquely identify above, should coincide with the one suggested in
\cite{Car99}. Our formulas (\ref{III.9}), (\ref{III.11}) in this case are
similar to formulas (7) from \cite{Car99}, but a non-trivial recalculation of
the Jacobian in the cochain  (\ref{III.10}) is required in order to bring it
into the form used in \cite{Car99}.

Comparison of our quantization formulas (\ref{III.9}), (\ref{III.11})
with the versions introduced in
\cite{UH78,Win84,Liu92,Lan93,BNW98,PFL98,BPW03} shows a difference in
the amplitude factor $k$ of (\ref{III.10}) and this corresponds with
the fact that the last two axioms in (\ref{II.6}) do not hold for
those other quantizations.

\section{Magneto-geodesic reflections}
\setcounter{equation}{0}

The magneto-geodesic quantization defined in the previous section is based on
the quantizer structure.  We now analyze this structure from the viewpoint of
symplectic transformations in the phase space $\prec\!\!\TB,\omega\!\!\succ$.

First note that the quantizer $\Delta_x$ can be decomposed into the product of
its unitary part $\Do_x$ and its modulus $|\Delta_x|$ as follows
\begin{equation}\label{IV.1}
\Delta_x = |\Delta_x|\,\cdot\Do_x\,,
\end{equation} where $x=(q,p) \in \TB$.
Here $|\Delta_x|$ is the positive square root of $\Delta_x^\dag
\Delta_x= \Delta_x^2$. It has the form of a multiplication
operator $|\Delta_x| = (\pi\h)^{-n}\sqrt j_q$.
The unitary part $\Do_x$ is
\begin{equation}\label{IV.2}
\Do_x = \sqrt{\frac {\D m(s_q)}{\D m} }\, \exp\left\{ {\frac
{2i}\h} p\,V_q + {\frac i\h} \Phi_q \right\}\, s_q^*\,,
\end{equation} where $s_q^*$ is the operator in $\H = L^2(\M, dm)$ generated by
the geodesic reflection
\begin{equation*}\label{IV.2a}
(s^*_q\psi)(q') = \psi\big(s_q(q')\big)\,, \qquad \psi \in \H\,.
\end{equation*}

    One can continue this decomposition and represent $\Do_x$
as the product of two unitary factors
\begin{equation}\label{IV.3}
\Do_x = E_x \cdot T_x \,,
\end{equation}
which are defined by


\begin{equation}\label{IV.4}
E_x  \equiv  \exp \left\{ {\frac {2i}\h} p\,V_q + {\frac i\h}
\Phi_q \right\}\,, \qquad T_x   \equiv  \sqrt{\frac {\D m(s_q)}{\D
m} } \, s_q^*\,.
\end{equation}
Note that both $E_x$ and $T_x$ are unitary
and they both commute with $|\Delta_x|$.
In addition, $T_x$ and $\Do_x$ are self-adjoint.

Now we would like to associate the unitary factors
in (\ref{IV.3}) with symplectic transformations of phase space. 
This is achieved by means of a Fock procedure like (\ref{III.33}).

Proceed first with the operator $T_x$. For each $x=(q,p)\in \TB$ define the
transformation $t_x$ of the phase space by the following formula:
\begin{equation}\label{IV.6}
t_x\left(\begin{array}{c} q' \\ p' \end{array}\right) =
\left(\begin{array}{l} s_q(q')\phantom{ \Big)}
\\ds_q(q')^{-1*}\big(p' + \beta_q(q')\big)
\end{array}\right)\,.
\end{equation}
Here $q'$ runs over $\M,\,\,  p'\in T^*_{q'}\M$, and the 1-form $\beta_q$ on
$\M$ is determined by
\begin{equation}\label{IV.7}
\beta_q = A^o - s_q^* A^o\,,
\end{equation}
where $A^o$ is the 1-form (\ref{III.22}) and  $s_q^*$ denotes the pullback of a
1-form. Obviously the mapping $t_x$ preserves the form $\omega$ (\ref{I.3}).
\begin{lem} \label{4.8} For any $g\in \P^1$ the permutation formula
\begin{equation}\label{IV.8}
T_x\, \g\, T_x^{-1} = \widehat{t_x^*\,g}
\end{equation} holds on the dense domain $\D(\M) \subset \H$.
\end{lem}

Formula (\ref{IV.8}) is easily derived from (\ref{III.21}) and (\ref{IV.6}). It
relates the unitary operator $T_x$ in the Hilbert space $\H = L^2(\M,dm)$ to
the symplectic transformation $t_x$ on the phase space
$\prec\!\!\TB,\omega\!\!\succ$.

Now we proceed in the same way with the operator $E_x$. For each $x=(q,p)\in
\TB$ let us define the transformation $e_x$ of phase space as follows
\begin{equation}\label{IV.9}
e_x\left(\begin{array}{c} q' \\ p' \end{array}\right) =
\left(\begin{array}{l} q'\phantom{ \Big)}
\\p' -2\, dV_q(q')^* p - d\Phi_q(q')
\end{array}\right)\,.
\end{equation}
Here $dV_q$ is the differential of the mapping
$V_q=\underline{\exp}_{\,q}^{-1}$. This transformation  preserves the form
$\omega$.

\begin{lem} \label{4.9} For any $g\in \P^1$ the permutation formula
\begin{equation}\label{IV.10}
E_x\, \g\, E_x^{-1} = \widehat{e_x^*\,g}
\end{equation} holds on the dense domain $\D(\M) \subset
\H$.

\end{lem}

This formula relates the unitary $E_x$ with the symplectic $e_x$ and follows
without difficulty from (\ref{III.21}) and (\ref{IV.4}).

As a consequence of these calculations we have
\begin{equation*}\label{IV.10a}
\Do_x\, \g\, {\Do_x^{-1}} = E_x\,T_x\, \g\, T_x^{-1} E_x^{-1} =
\widehat{e_x^*\,t_x^*\, g}\,,
\end{equation*}
for any $g\in \P^1$.  Thus one obtains the composition of two symplectic
transformations
\begin{equation}\label{IV.11}
\sigma_x \df t_x\circ e_x\,.
\end{equation}

\begin{cor}\label{cor2} The symplectic transformations
$\{\sigma_x\}$
are related to the unitary part of the quantizer by the identity
\begin{equation}\label{IV.12}
\Do_x\, \widehat g\, \Do_x^{-1} = \widehat{\sigma_x^*\, g}\,,
\end{equation}
for any $g\in \P^1$, and any $x\in \TB$. These identities hold on
the dense domain $\D(\M) \subset \H$\,.
\end{cor}

From formulas (\ref{IV.6}), (\ref{IV.9}) and from the equalities
\begin{equation}\label{IV.13}
d\Phi_q(q') = \beta_q(q') - \alpha_q(q')\,, \qquad \alpha_q(q') \df
A\big(q',q\big) - ds_q(q')^*A(s_q(q'),q)\,,
\end{equation}
the symplectic mapping $\sigma_x$ is determined to be
\begin{equation}\label{IV.14}
\sigma_x\left(\begin{array}{c} q' \\ p' \end{array}\right) =
\left(\begin{array}{l} s_q(q')\phantom{ \Big)}
\\ds_q(q')^{-1*}\big[p' - 2\, dV_q(q')^*p + \alpha_q(q')\big]
\end{array}\right)\,.
\end{equation}
Here the potential $A$ is given by (\ref{III.32}). Note the potential $A$ and
thereby the right hand side above is gauge independent although (\ref{IV.6})
and (\ref{IV.9}) are gauge dependent (they depend on a choice of the point $o$
in the potential $A^o$ in (\ref{III.22})).

Let us check the simplest properties of the family of mappings
(\ref{IV.14}). If the running point $x' =(q',p')$ coincides with
$x$, then $\sigma_x(x')|_{x'=x} = x$. Thus the point $x$ is a
fixed point of the transformation $\sigma_x$.

Furthermore, from the evident quantum permutation relations
\begin{equation*}\label{IV.32a}
 T_x\,E_x = E_x^{-1}\, T_x\,, \qquad T_x^2 = I
\end{equation*} one obtains the corresponding classical counterparts
\begin{equation*}\label{IV.32b}
e_x\circ t_x = t_x \circ e_x^{-1}\,, \qquad t_x^2 = id\,.
\end{equation*}
The definition (\ref{IV.11}) then implies
\begin{equation*}\label{IV.32c}
\sigma_x^2 = t_x\circ e_x \circ t_x \circ e_x = t_x \circ t_x
\circ e_x^{-1} \circ e_x = id\,.
\end{equation*} Thus Corollary {\ref{cor2}} can be completed as follows.

\begin{cor}\label{cor3} The family $\{\sigma_x\,|\, x\in \TB\}$ of
symplectic transformations on the space $\prec\!\!\TB,\omega\!\!\succ$ is given
by formula {\rm{(\ref{IV.14})}} and possesses the properties{\rm{:}}
\begin{itemize}
  \item $x$ is a unique and isolated fixed point of $\sigma_x$\,,
  \item each $\sigma_x$ is a reflection, \ie $\sigma_x^2 = id$\,.
\end{itemize}
\end{cor}

This family of reflections can be considered as a lift to $\TB$ of
the family of reflections $\{s_x\}$ given on $\M$. We call
$\sigma_x$ a {\it magneto-geodesic reflection}. The maps
$\{\sigma_x\}$  generalize the family of magnetic reflections
found in \cite{KO3} for the phase space
$\prec\!\!T^*\R^n,\omega\!\!\succ$ with the Euclidean connection
on $\M=\R^n$. Now our reflections (\ref{IV.14}) combine both: a
nontrivial magnetic field and a nontrivial affine connection on
$\M$.

We note that the correspondence between the quantizer and the
phase space reflections was observed  in \cite{Gro76} for the
Euclidean space $\R^{2n} = T^*\R^n$ with no magnetic field, see
also \cite{Roy77,GH78}. In this latter case the reflections are
just $\sigma_x(x') = 2x - x'\,.$

In the general phase space $\TB$ by using the magneto-geodesic reflections
$\sigma_x$ one can easily define the notions of $\sigma$-midpoints and
$\sigma$-reflective curves.

Namely, the point $x\in \TB$ is called a $\sigma$-midpoint between $x',x'' \in
\TB$ if $x''= \sigma_x(x')$. From (\ref{IV.14}) it follows that any two points
from $\TB$ have a unique $\sigma$-midpoint.

A continuous curve passing through a point $x \in T^*\M$ is called
$\sigma$-{\it reflective with respect to $x$} if it consists of pairs of points
with the midpoint $x$. The projection of the $\sigma$-reflective curve from
$\TB$ onto $\M$ is a reflective curve in $\M$, and vice versa, any reflective
curve from $\M$ (in particular, the geodesic) can be lifted to a
$\sigma$-reflective curve on $\TB$ (but of course, not uniquely).

\section{Integral formula for quantum product}
\setcounter{equation}{0}

Next we present an explicit formula for the product $f\star g$, expressed
directly in terms of the functions $f$ and $g$. The definition of $\star$
product was given in (\ref{III.16}). In principle, one could use (\ref{III.16})
to compute $f\star g$. But it is convenient to use the equivalent
representation (\ref{II.7}).

First we compute the distribution (\ref{II.7a}), that is the trace
of the composition of three quantizers $\Delta_x\Delta_y\Delta_z$.
Label the phase space points here by
\begin{equation}\label{V.0}
x=(a,\xi),\,  y=(b,\eta),\,  z=(c,\zeta), \quad {\mathrm{ where}}\,\, \xi \in
T_a^*\M,\, \eta \in T_b^*\M,  \zeta \in T_c^*\M.
\end{equation}
Let $(q,q')$ denote the two arguments of the integral kernel of the composition
$\Delta_x\Delta_y\Delta_z$. Employing (\ref{III.3}) gives
\begin{eqnarray}\label{V.1}
&\N\cdot {\mathrm{Kernel}}\big(\Delta_x\Delta_y\Delta_z\big)(q,q')
=
T_{a,b,c}^{\xi,\eta,\zeta}(q)\,\delta_{s_c s_b s_a(q)}(q')\,,\phantom{\bigg)} \\
\label{V.2}&T_{a,b,c}^{\xi,\eta,\zeta}(q) = (\pi \h)^{-2n} \exp\left\{
S_{a,b,c}^{\xi,\eta,\zeta}\right\}\, \varphi_{a,b,c}(q)\,.\qquad \, \quad
\end{eqnarray}
The phase and amplitude functions in (\ref{V.2}) are
\begin{equation}\label{V.3}
S_{a,b,c}^{\xi,\eta,\zeta}(q) = 2\left[ \xi V_a(q) +\eta V_b\big(s_a(q)\big)
 + \zeta V_c\big(s_b s_a(q)\big)\right]\,
+ \Phi_a(q) + \Phi_b(s_a(q)) + \Phi_c (s_b s_a(q))\,, \end{equation}
\begin{equation}
\label{V.4} \varphi_{a,b,c}(q) = 2^n\,\big[J_a(q)\,J_b(s_a(q))\,
J_c(s_bs_a(q))\big]^{1/2}\,.\qquad \, \quad \phantom{\bigg)}
\end{equation}
Clearly, $T_{a,b,c}^{\xi,\eta,\zeta}(q)$ is a non-singular, continuous function
of all its arguments.

The integral kernel (\ref{V.1}) is singular. Therefore, in order
to evaluate the trace of the corresponding operator, we first
contract the distribution (\ref{V.1}) with a test function $\phi
\in \D(\M)$ by  the parameter $c\in \M$. This computation is
implemented by using the formula
\begin{equation}\label{V.5}
\int_{\M}\delta_{s_c s_b s_a(q)}(q')\,\phi(c)\,dm(c) = \phi(c)\,{
\bigg |\frac {\D m(s_c s_b s_a (q))}{\D m(c)} }\bigg
|^{-1}_{\,c=q'\vee s_b s_a(q)}\,.
\end{equation}
On the right hand side of this formula the point $c$ is taken to be the
mid-point $q'\vee s_b s_a(q)$, (see Fig. 4).

\begin{figure}
\centering
\includegraphics{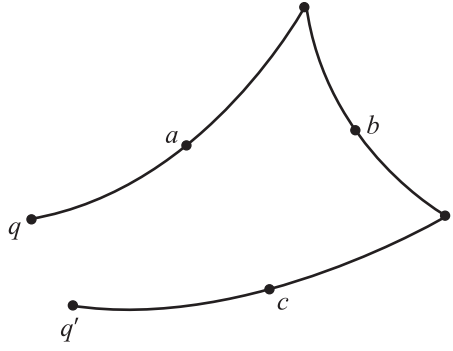}
\caption{Mapping $s_c s_b s_a(q) = q'$. }
\end{figure}

In order to evaluate the trace we have to put $q=q'$ and integrate
over all $q\in \M$. Thus from (\ref{V.1}), (\ref{V.5}) one obtains
\begin{equation}\label{V.6}
\N\! \int_{\M}\! \Tr \left
(\Delta_{(a,\xi)}\,\Delta_{(b,\eta)}\,\Delta_{(c,\zeta)}\right )\,
\phi(c)\,dm(c) = \int_{\M}\! T_{a,b,c}^{\xi,\eta,\zeta}(q)\frac{\phi(c)}{\big
|{\frac {\D m(s_c s_b s_a (q))}{\D m(c)}\big | } } \bigg |_{\,c=q\vee s_b s_a
(q)}\, dm(q)\,.
\end{equation}

 The mapping
 \begin{equation}\label{V.7} q\mapsto c=q\vee s_b s_a
(q) \end{equation} is smooth, but it can be degenerate. The Jacobian of this
mapping is
\begin{equation}\label{V.8}
\left | \frac{\D m(c)}{\D m(q)} \right | = \big | \det\big(I - d(s_c s_b
s_a)(q)\big) \big |\cdot \left | \frac{\D m(s_c s_b s_a (q))}{\D m(c)} \right
|^{-1}\,.
\end{equation}
So, the degeneracy of (\ref{V.7}) is controlled by the determinant
\begin{equation}\label{V.9}
\J_{a,b,c} (q) \df \big | \det\big(I - d(s_c s_b s_a)(q)\big) \big |\,.
\end{equation}
Here the point $c$ is assumed to be the image point of the mapping (\ref{V.7})

In each connected domain $\M_0 \subset \M$, where the Jacobian (\ref{V.9}) is
not zero, one can invert the mapping (\ref{V.7}) and uniquely express $q$ as a
function of $c$ (and of $a,b$ as well): $q = Q(a,b,c)$. Obviously $Q$ is the
fixed point of the composition of three reflections (see Fig. 5):
\begin{equation}\label{V.10}
s_c s_b s_a (Q) = Q\,.
\end{equation}
Inside the domain $\M_0$ the solution of this fixed point problem exists and is
unique. At the boundary of the domain $\M_0$, where the Jacobian (\ref{V.9})
becomes zero, the solution of (\ref{V.10}) is not infinitesimally isolated.

\begin{figure}
\centering
\includegraphics{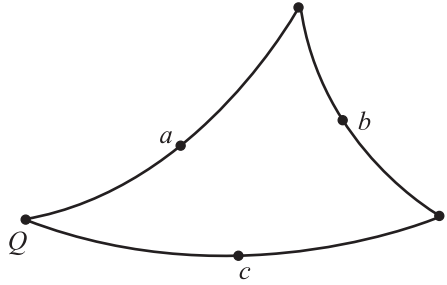}
\caption{Fixed point $Q$ of the compound reflection $s_c s_b s_a$.}
\end{figure}

From (\ref{V.6}), (\ref{V.8}) it follows that
\begin{equation*}\
\N\! \int_{\M}\! \Tr \left
(\Delta_{(a,\xi)}\,\Delta_{(b,\eta)}\,\Delta_{(c,\zeta)}\right )
\phi(c)\,dm(c) = \int_{\M}\! T_{a,b,c}^{\xi,\eta,\zeta}(q)
\J_{a,b,c}(q)^{-1}\phi(c) \left |\frac {\D m(c)}{\D (q)}\right |
\, dm(q)\,,
\end{equation*}
and so, one has the following formula for the distribution $K_\star$ in
(\ref{II.7a})
\begin{equation}\label{V.11}
K_\star(x,y,z) = \N\, \Tr \left
(\Delta_{(a,\xi)}\,\Delta_{(b,\eta)}\,\Delta_{(c,\zeta)}\right ) =
\sum_{Q} \frac 1{\J_{a,b,c}(Q)}\ T_{a,b,c}^{\xi,\eta,\zeta}(Q)\,.
\end{equation}
The summation is taken over all fixed points of (\ref{V.10}).

If the triple $(a,b,c)$ is such that there is no solution of the fixed point
problem (\ref{V.10}), then the value of the trace (\ref{V.11}) is just zero.

If the triple $(a,b,c)$ is such that there is a solution of the fixed point
problem (\ref{V.10}) which is not infinitesimally isolated (the Jacobian
(\ref{V.9}) is zero), then the value of the trace (\ref{V.11}) is infinite. The
precise description of what this `infinity' actually is, is given by the
integral (\ref{V.6}) (where there are no singularities at all).

Note that the amplitude factor on the right hand side of formula
(\ref{V.11}) is derived from (\ref{V.2}), (\ref{V.4}) and the
definition of the Jacobian $J_a$ at the beginning of Sect.~3.
Thus we obtain the amplitude
\begin{equation}\label{V.12}
\varphi(a,b,c)\df
\varphi_{a,b,c}(Q) = 2^n \left[ j_a(Q)j_b(s_a(Q)j_c(s_b s_a(Q))\cdot
\big | \det d(s_c s_b s_a)(Q) \big | \right ]^{1/2}\,,
\end{equation}
where $Q=Q(a,b,c)$ is the solution of (\ref{V.10}).

The phase of the exponential factor in (\ref{V.11}) can be
represented in the following geometrical form.
\begin{lem}\label{lem10}
\begin{equation}\label{V.13}
S_{a,b,c}^{\xi,\eta,\zeta}(Q) = \int_{\Sigma(x,y,z)} \omega\,.
\end{equation}
Here the symplectic form $\omega$ is determined by {\rm(\ref{I.3})}, the points
$ x,y,z \in \TB$ are given by {\rm(\ref{V.0})} and $\Sigma(x,y,z)$ is a
triangle in $\TB$ whose sides are $\sigma$-reflective curves with midpoints
$x,y,z$.
\end{lem}

Indeed, the symplectic area on the right hand side of (\ref{V.13}), by Stokes
theorem, can be represented as the sum of three integrals
\begin{equation}\label{V.14}
\int_{\Sigma(x,y,z)} \omega = \int_{Q'}^Q (\tilde{p}\,d\tilde q + A^0) +
\int_{Q''}^{Q'} (\tilde{p}\,d\tilde q + A^0) + \int_{Q}^{Q''}
(\tilde{p}\,d\tilde q + A^0)\,.
\end{equation}
Here $Q'=s_a(Q)\,, Q'' = s_b s_a(Q)\,,$ each integral (\ref{V.14}) is taken
along the geodesics through the midpoints $a,b,c$ respectively, and $\tilde p
\in \TB$ denotes the value of the momentum on the $\sigma$-reflective curves
over these geodesics. The magnetic potential $A^0$ is a primitive of the
Faraday form $dA^0 = F$. Using (\ref{III.21b}) we conclude that the three
integrals of $A^0$ in (\ref{V.14}) correspond to three summands with the
functions $\Phi_a, \Phi_b, \Phi_c$ in (\ref{V.3}), and the three other
integrals of $\tilde{p}\,d\tilde q$ contribute to the terms containing the
momenta $\xi,\eta,\zeta$. For example, in the first integral of (\ref{V.14})
one has
\begin{equation*}
\int_{Q'}^Q \tilde{p}\,d\tilde q = \int_{Q'}^Q \big(dV_a(\tilde q)^*\xi -
b_a(\tilde q)\big)\, d\tilde q = \int_{Q'}^Q d\big(\xi V_a(\tilde q)\big) = \xi
V_a(Q) -\xi V_a(Q') = 2\xi V_a(Q) \,.
\end{equation*}
In the above formula $b_a\equiv A^0 - \frac 12 d\Phi_a$ is used to make the
curve $\{(\tilde q, \tilde p)\}$ to be $\sigma$-reflective with respect to the
midpoint $x=(a,\xi)$.

Thus from (\ref{II.7}), (\ref{II.7a}) and the formulas
(\ref{V.11}), (\ref{V.12}), (\ref{V.13}) we obtain the following
result

\begin{equation}\label{I.5}
(f \star g )(x) = \frac 1{(\pi\h)^{2n}}\int\!\!\int_{C(\xb)}\sum  \exp\Big\{
\frac i \hbar \int_{\Sigma(x, y, z)}\omega \Big\}\,
\frac{\varphi(\xb,\yb,\zb)}{ \J(\xb,\yb,\zb)} f(y) g(z)\,dy \,d z\,.
\end{equation}
In this formula
\begin{itemize}
  \item all objects are independent of the choice of the measure
  on the manifold $\M$ and are determined by an affine structure
  $\underline{\Gamma}$ and a closed $2$-form $F$ on $\M$;
  \item points $x,y,z$ belong to the phase space $\TB$, and points
  $\xb,\yb,\zb$ are their projections onto $\M$;
  \item the positive function $\varphi\in C^\infty(\M\times\M\times\M)$
  is given by (\ref{V.12});
  \item the Jacobian $\J(\xb,\yb,\zb) = \big|\det\big( I -
  d(s_{\zb}s_{\yb}s_{\xb})(Q)\big)\big|$
  is determined  by the differential of the composition
  $s_{\zb}s_{\yb}s_{\xb}$ of three $\underline{\Gamma}$-geodesic
  reflections in $\M$ with respect to
  midpoints $\zb,\yb,\xb $, and where $Q$ is the fixed point of this
  composition;
  \item the magnetic form $\omega$ (\ref{I.3}) is integrated in (\ref{I.5})
  over surfaces (or membranes) which are `triangles' $\Sigma(x,y,z)$ in $\TB$
  whose sides are $\sigma$-reflective curves with midpoints
$x,y,z$. The projection of $\Sigma(x,y,z)$  onto $\M$ are geodesic
triangles $\underline \Sigma(\xb,\yb,\zb)$ with midpoints
$\xb,\yb,\zb$;
 \item the domain of integration in (\ref{I.5}) is a cotangent `cylinder'
$C(\xb) = T^*(\MM)$ over the subset $\MM$ consisting of all those
pairs of points $\yb,\zb \in \M\times\M$ for which the triangle
$\underline \Sigma(\xb,\yb,\zb)$ exists. The sum $\sum$ is taken
over all such triangles.
\end{itemize}

\begin{theor} \label{thm1} The associative product of functions over the phase space $\TB$,
which corresponds via {\rm(\ref{II.3})} to the magneto-geodesic
quantization, is determined by the formula {\rm(\ref{I.5})}. This
formula acts directly on the subalgebra $\SF(\TB)$ of functions
whose momentum Fourier image belongs to $\D(T\M)$, and is extended
to the algebra $\SF_{\star}$
{\rm(}and to its subalgebra $\P${\rm)}
by the procedure {\rm{(\ref{II.10})}}.
\end{theor}

The subset $\MM$ can be called the {\emph{domain of influence}} of the point
$\xb \in\M$. Inside $\MM$ the Jacobian\! $\J$ is not zero. In general the
domain of influence $\MM$ does not coincide with the whole $\M\times\M$. So, if
the function $f\otimes g$ is localized outside of $C(\xb)$, then the integral
(\ref{I.5}) vanishes in a neighborhood of the fiber $T^*_{\xb}\M$.

As an example consider the Lobachevski plane, $\M=H^2$, given by
the hyperboloid in three dimensional Euclidean space : $H^2
\equiv\{q\in \R^3 |\, q_1^2 + q_2^2 - q_3^2 = -1\}$. The
Riemannian structure on $H^2$ is induced from the Euclidean
structure on $\R^3$; the connection $\underline{\nabla}$ on $H^2$
is the usual Levi-Civita connection. In this case it is known (\cf
\cite{Tuy99,Rios02} ) that the geodesic triangle with midpoints
$\underline{x}\,\underline{y}\,\underline{z}$ exists iff
\begin{equation}\label{V.15}
-1 < \det|\underline{x}\,\underline{y}\,\underline{z}| < 1
\end{equation}
(here $\underline{x}\,\underline{y}\,\underline{z}$ are considered to be
3-vectors, and $|\underline{x}\,\underline{y}\,\underline{z}|$ denotes the
$3\times 3$ matrix of their components). Under the condition (\ref{V.15}) such
a triangle is unique. If one fixes $\underline{y}\,,\underline{z}$ then the
subset $H_{\underline{y}\,\underline{z}}$ of points $\underline{x}\in H^2$
obeying (\ref{V.15}) looks like a tubular neighborhood of the geodesic passing
through $b$ and $c$. Certainly $H_{\underline{y}\,\underline{z}}$ is a proper
subset in $H^2$, that is $H_{\underline{y}\,\underline{z}} \neq H^2$ if
$\underline{y} \neq \underline{z}$. Thus, if $\underline{x} \notin
H_{\underline{y}\,\underline{z}}$ then $(\underline{y}\,,\underline{z}) \notin
H^2\times_{\xb} H^2$. Therefore, the domain of influence $H^2 \times_{\xb} H^2$
is a proper subset in $ H^2 \times H^2$.

We see that the quantum product (\ref{I.5}) determines the
distribution
\begin{equation}\label{I.7a}
K_\star(f\otimes g\otimes l) = \langle f\star g, l \rangle\,,
\end{equation}
whose support, in general, does not coincide with the whole
$\TB\times\TB\times\TB$. The topological boundary of this support
can be considered as a {\emph{quantum front}}
\begin{equation}\label{I.7b}
{\mathrm {front}}\,(\star) \df \pr({\mathrm {supp}}\, K_\star)\,.
\end{equation}
From one side of this boundary the distribution $K_\star$ is
identically zero. We call this phenomenon a {\emph{front-effect}}.

Something close to this was mentioned in the interesting note
\cite{Rios02} following ideas of geometric quantization, but no
actual construction of any associative product was produced there.

The phase space of the type $\TB$, which we investigate in this
paper, and in particular the space $T^*H^2$ seems to be the first
instance where the front-effect for the $\star$ product is
mathematically identified.

The kernel $K_\star$ can be considered as a product of
two $\delta$-functions
\begin{equation*}
K_\star(x,y,z) = \big(\delta_y \star \delta_z\big)(x)\,.
\end{equation*}
The set supp$(\delta_y \star \delta_z)$ is the cylinder
$T^*\M_{\underline{y}\underline{z}}$ over the domain
$\M_{\underline{y}\underline{z}}\subset \M\times \M$ (consisting
of those $\underline{x}$ for which the geodesic triangle
$\underline{\Sigma}(\underline{x},\underline{y},\underline{z})$
exists). Outside of this cylinder the distribution $\delta_y \star
\delta_z$ is identically zero. The boundary of
$T^*\M_{\underline{y}\underline{z}}$ is a front of the `wave'
$\delta_y \star \delta_z$ which travels in the phase space $\TB$
when the points $\underline{y}$ and $\underline{z}$ move away from
each other.

Note that it is not possible to detect the front-effect in
classical mechanics or even in  the formal deformation approach
where instead of exact associative product like (\ref{I.5}) one
uses a formal asymptotic power series in $\h$.

Also note that in the approach based on ideas of pseudodifferential operator
theory \cite{Lan93,BNW98,BNW98b,Wid80}, where the $\star$-product is considered
only on symbols whose $p$-Fourier transform is localized near zero, one can
introduce under the integral (\ref{I.5}) a cutoff function.  This function is
identically 1 for $x,y,z$ close enough to each other, but becomes $0$ if
$x,y,z$ move away from each other. In this way one can eliminate all the
difficulties related to the possible existence of conjugate points on geodesics
or the non-existence of triangles. The formula (\ref{I.5}), with the cutoff
function, works for arbitrary affine manifolds $\M$;
no front effect exists in
this approach and no summation over multiple triangles is needed.
Of course,
this `cutoff method,' in general,  destroys the associativity of
the $\star$-product.
But with this approach one can keep associativity in the
algebra $\P$ of symbols polynomial in momenta.
Thus, we conclude that {\it formula {\rm(\ref{I.5})} works in
the algebra $\P$ over an arbitrary manifold~$\M$}.

Looking forward to Sect.~6 we note that the coefficients of the
asymptotic expansion (\ref{I.6}), (\ref{VI.1}) of the $\star$ product
are insensitive to the cutoff function, since they are derived from
the diagonal of the exact product (\ref{I.5}). Thus, the $\hbar\to0$
asymptotic expansion of the  product (\ref{I.5}) works over an
arbitrary manifold $\M$.

We stress that the core result of Theorem~1 is the explicit formula for an
associative quantum product over the phase space $T^*{\cal M}$. This formula is
exact, not a deformation one, and even not a semiclassical one. Under the
semiclassical approach the leading term of the asymptotics of the product
kernel $K_\star$ is known over general phase spaces \cite{new-a}.
The membrane formula like (\ref{V.13}) for the
phase of $K_\star$ was
first suggested in \cite{new-b}
as a formula for the `action' on the graph of groupoid multiplication,
and it was proved in \cite{new-a},
in the general symplectic case,
that indeed (\ref{V.13}) is the correct solution of the Cauchy
problem for the phase function.
In the Euclidean case, membrane
formulas of a similar type were discovered by M.~Berry \cite{new-c} for the
asymptotics of the Wigner function (see also \cite{new-d} for solutions of the
Cauchy problem). The magnetic version of these formulas was first obtained and
investigated in detail in \cite{KO1,KO3}, the case of symmetric spaces was
studied in \cite{new-e}, and for general manifolds, see in \cite{new-f,new-g}.

\section{Magneto-geodesic connection and $\h$-expansion of the quantum product}
\setcounter{equation}{0}
\setcounter{cor}{4}

 In this section we investigate the asymptotic properties of
the product $f\star g$ in the classical approximation
as $\h \rightarrow 0$ and express the
results in terms of magneto-geodesic covariant derivatives over
phase space $\TB$. It is assumed that the functions $f,g$ are
$\h$-independent.

The integral (\ref{I.5}) determing  the quantum product contains a
rapidly oscillating exponential factor and a smooth (non-oscillating)
amplitude. The exponent phase has a stationary point at $y=x,\, z=x$,
which is  isolated and non-degenerate. Therefore one can apply the
standard stationary phase method \cite{Fed71} to derive the
asymptotic expansion of the integral (\ref{I.5}) as $\h\rightarrow
0$. The structure of this expansion is
\begin{equation}\label{VI.1}
    (f\star g)(x) = \sum_{k\geq 0} \frac 1{k!}\left( {\frac{-i\h}2} \right)^k G_k(f\otimes
    g)(x)\,.
\end{equation}
Here $G_k$ are differential operators of order $2k$ acting on the
function $f(y)\,g(z)$ and then restricted to the diagonal (the
stationary point) $y=x,\, z=x\,$.

 Since we know that the unity function $1$ is
the unit element for the $\star$-product, then
\begin{equation} \label{VI.2}
\begin{array}{l} G_0(f\otimes 1) = f\,, \phantom{\Big (} \\
G_k(f\otimes 1) = G_k(1\otimes f) = 0\,, \qquad k\geq  1\,. \end{array}
\end{equation}
From (\ref{VI.2}) it follows, for instance, that
\begin{equation*} \label{VI.2a}
\quad G_0(f\otimes g) = fg\,, \qquad G_1(f\otimes g) = \langle df, \Psi\, dg
\rangle
\end{equation*}
where $\Psi$ is a 2-tensor. From the property (\ref{II.7b}) and
the involution property $\overline{f \star g} =
    \overline{g}\star \overline{f}$
we see that $\Psi$ must be skew-symmetric and real. In addition, the
associativity of the $\star$-product implies the Jacobi identity for $\Psi$.
Thus $\Psi$ is a Poisson tensor on $\TB$. It is easy to check that $\Psi =
\Omega^{-1}$, where $\Omega$ is the matrix of the symplectic form $\omega$ in
the exponent of (\ref{I.5}).

Thus the expansion of (\ref{VI.1}) takes the form
\begin{equation}\label{VI.3}
    f\star g = fg - \frac{i\h}2 \{f,g\} - \frac{\h^2}8\, G_2(f\otimes g)\big |_{diag}
    + O(\h^3)
\end{equation}
with the Poisson bracket $\{f,g\} \equiv \langle df, \Psi dg \rangle \,.$

In the general theory of deformation quantization
\cite{BBF78,OMY91,Fed94,GR03} it was observed that the operator
$G_2$ in expansions of star-products like (\ref{VI.3}) has to be
related to a certain phase space connection $\nabla$, and
moreover, $G_2$ can be written in the form
\begin{equation}\label{VI.4}
    G_2 = (\nabla'\Psi \nabla'')^2 + \nabla' \B \nabla''\,.
\end{equation}
Here $\B$ is a certain 2-tensor, and the primes mark the argument (first or
second) of the function $f\otimes g$ to which the covariant derivative $\nabla$
is applied.

For our specific (and exact) quantum product $\star$ over $\TB$
the generic form $(\ref{VI.4})$ can be verified and explicit
formulas for $\nabla$ and $\B$ can be found directly from the
asymptotic expansion of the integral (\ref{I.5}).

But we also could match the quantization with a phase space connection in
another way. The quantization is given by the quantizer $\Delta_x$, which in
turn defines a family of symplectic transformations $\sigma_x$ via
(\ref{IV.14}). The transformations $\sigma_x$ acting on $\TB$ are the phase
space analogs of the geodesic reflections $s_q$ acting on $\M$. These later
reflections are generated by the connection $\underline{\nabla}$ over $\M$. The
Christoffel symbols of this connection are determined via $s_q$ as follows
\begin{equation*}\
\underline{\Gamma}(q) = - \frac 12\, D^2 s_q(q'){\big|}_{q'=q}\,,
\end{equation*}
where $D$ denotes derivatives with respect to argument $q'\in \M$. We  can just
mimic this formula to define a connection over $\TB$
\begin{equation}\label{VI.5}
{\Gamma}(x) = - \frac 12\, D^2 \sigma_x(x'){\big|}_{x'=x}\,,
\end{equation}
where $D$ denotes derivatives by the argument $x'\in\TB$. This
formula indeed generates a connection for any family of reflections,
and such a connection is automatically symplectic if these
reflections are symplectic~\cite{new-a}. In our magneto-geodesic
situation we have all these properties of the family $\{\sigma_x\}$
(Corollary~\ref{cor3}; for the Euclidean case $\M = R^n$, see more
details in~\cite{KO3}).

Note that the restriction of $\sigma_x$ to the configuration space $\M$
coincides with $s_q$ (see (\ref{IV.14})), and so, the set of Christoffel
symbols $\Gamma$ (\ref{VI.5}) contains the Christoffel symbols
$\underline{\Gamma}$ inside itself. Thus $\Gamma$ is an extension of
$\underline{\Gamma}$ from the configuration space to the phase space.

Explicit calculation of the second derivatives in (\ref{VI.5}) using
(\ref{IV.14}), yields
\begin{equation} \label{VI.5a}
 \Gamma_{qq}^q = {\underline{\Gamma}}(q)\,, \qquad  \Gamma_{qp}^p
=\Gamma_{pq}^p = -\underline{\Gamma}(\,q)\,, \qquad \Gamma_{qq}^p = p\, B(q) +
C(q)\,, \tag{\ref{VI.5}$a$}
\end{equation}
where
\begin{equation} \label{VI.5b}
B^m_{jkl} \df \frac 13 \, \underset{jkl}{\mathfrak S} \left(2\,
\underline{\Gamma}_{\,js}^m\, \underline{\Gamma}_{\, kl}^s -
\pr_j\, \underline{\Gamma}_{\,kl}^m\right)\,, \qquad C_{jkl} \df
\frac 13 \,(\underline{\nabla}_{\,k}\, F_{jl} +
\underline{\nabla}_{\,l}\, F_{jk})\,, \tag{\ref{VI.5}$b$}
\end{equation}
and the notation ${\mathfrak S}$ denotes cyclic summation. All other components
of $\Gamma$ vanish identically: $\Gamma_{pq}^q = \Gamma_{qp}^q = \Gamma_{pp}^q
= \Gamma_{pp}^p = 0$\,. \smallskip

Let $\nabla$ be the covariant derivative over $\TB$ defined by
Christoffel symbols (\ref{VI.5a}). We call it the
{\emph{magneto-geodesic connection}}. We stress again that this
connection is symplectic with respect to  the `magnetic' symplectic
structure $\omega$ (\ref{I.3}), i.e.,
$$
\nabla \omega = 0.
$$

We see that such a $\nabla$ is certainly related to the
quantizer and so also to
the $\star$-product (\ref{I.5}). We claim that this is actually the same
connection which appears in (\ref{VI.4}) under the $\h$-expansion of integral
(\ref{I.5}).

Now we demonstrate how to prove this claim. Let us fix $x=(a,\xi)\in \TB$.
Formula (\ref{V.3}) implies that there is just one stationary point of the
phase function $S=S^{\xi,\eta,\zeta}_{a,b,c}$ with respect to variables
$y=(b,\eta)$ and $z=(c,\zeta)$, namely, this is the point $y=x\,, z=x$. Near
this point only one summand in (\ref{I.5}) contributes to the asymptotic
expansion up to $O(\h^\infty)$, namely, this is the summand corresponding to
small triangles $\Sigma = \Sigma(x,y,z)$ near the point $x$. Consider $x$ as
the origin of the normal coordinate system (with respect to the connection
(\ref{VI.5})). We employ normal coordinates for  both variables $y,z$ in
(\ref{I.5}).  With this adjustment the integral looks like
\begin{equation}\label{VI.8}
    f\star g \sim \int_{\R^{2n}}\int_{\R^{2n}}\,e^{{\frac
    i\h}S(v',v'')} \,L(v',v'')\,f\big(\exp_x(v')\big)\,g\big(\exp_x(v'')\big)\, {\frac
    {dv'\,dv''}{(\pi\h)^{2n}}}\,.
\end{equation}
Here the integration space, $\R^{2n}\times\R^{2n}$, is just $T_x(\TB)\times
T_x(\TB)$; $v'\, v''$ are normal coordinates, and the phase function $S =
\int_\Sigma \omega$, \cf (\ref{V.13}). The amplitude function $L$ is given by
(\ref{V.9}), (\ref{V.12}): namely, $L=(\varphi/\J)\,l\otimes l$, where $l$ is
the density of the Liouville measure on $\TB$ expressed in normal coordinates.

\begin{lem}\label {lem10a} The following Taylor expansions hold
\begin{equation}\label{VI.9}
S = S_2 + S_4 + O^6\,, \quad L = 1 + L_2 + O^4\,, \quad l= 1+l_2 + O^4
\end{equation}
where $S_2, L_2, l_2$ are homogeneous polynomials of degree $2$, and $S_4$ is
of degree $4$; the remainders $O^4\,, O^6$ are of degree $4$ and $6$ in normal
coordinates near zero in $\R^{2n}\otimes \R^{2n}$. Formulas for the quadratic
forms $S_2$ and $L_2$ are
\begin{equation}\label{VI.10}
  S_2(v',v'')=  2 \, \langle\Omega v'', v'\rangle\,, \qquad L_2(v',v'') = -{\frac
  12}\, \langle \underline{\RR}(\underline{v}'-\underline{v}''),
  \underline{v}'-\underline{v}''
  \rangle + l_2(v') + l_2(v'')\,.
\end{equation}
Here $\Omega$ is the $2n\times2n$ matrix of the symplectic form,
$\Omega=\big[\begin{array}{cc}F&I\\-I&0\end{array}\big]$, $\underline{v}\in
\R^n$ denotes the $q$-components of the vector $v\in \R^{2n} = \R_q^n\times
\R_p^n$, and $\underline{\RR}$ is the symmetric part of the $n\times n$ Ricci
tensor on $\M$
\begin{equation}\label{VI.11}
 \underline{\RR}_{\, ls} \df {\frac 12}\, (\underline{R}^{\,k}_{\ lks} +
 \underline{R}^{\,k}_{\ skl})\,, \end{equation}
where $\underline{R} = [\,\underline{\nabla}\,,
 \underline{\nabla}\,]$ is the curvature tensor (skew symmetric in the last
 pair of indices).
\end{lem}

\noindent {\it Proof.}  The first formula in (\ref{VI.10}) is obvious. Indeed
from (\ref{V.13}) we claim that
\begin{equation*}
    S = \frac 12\, \Omega\, V''\, V' + O^4\,,
\end{equation*}
where $V',V''$ are the sides of the linearization of the triangle
$\Sigma(x,y,z)$ at the point $x=(a,\xi)$. The sides $V',V''$ are twice as long
as the normal coordinates $v',v''$ of the midpoints $y=(b,\eta), z=(c,\zeta)$
of this triangle. This is the reason for the factor of $2$ in the first formula
of (\ref{VI.10}).

To prove the second formula in (\ref{VI.10}) we first deduce from (\ref{V.9})
and (\ref{V.12}) that
\begin{equation}\label{VI.12}
    L_2(v',v'') = l_2(v') + l_2(v'') + \alpha_2(\underline{v}',\underline{v}'') + \beta_2(\underline{v}',\underline{v}'')\,.
\end{equation}
Here $\alpha_2$ and $\beta_2$ are the second order Taylor terms about the point
$a$ of the functions
\begin{equation}\label{VI.14}\begin{array}{l}
    \alpha \df 2^n \big|\det d(s_c s_b s_a)(Q)\big|^{1/2}\,\J^{-1}(Q) \,,\phantom{\Biggr)}\\
    \beta \df \big[j_a(Q)\,j_b(s_a(Q))\,j_c(s_bs_a(Q))\big]^{1/2}
\end{array}\end{equation}
where $Q$ is the solution of the problem (\ref{V.10}) in the neighborhood of
the origin point $a$.

The Taylor expansion of $Q$, in normal coordinates, results from
\begin{equation}\label{VI.15}
    Q= \underline{\exp}\,_{a}(\underline{v}'' -\underline{v}' +O^2)\,.
\end{equation}
The Jacobi matrix of the mapping $s_c s_bs_a$ reads
\begin{equation*}
    d(s_c s_b s_a)\big(\underline{\exp}\,_{a}(\underline{v})\big) = -I +2\,(\underline{v}''-\underline{v}'
    -\underline{v})\,\underline{\Gamma}+ O^2\,.
\end{equation*}
From here and (\ref{VI.15}) it follows that
\begin{equation*}
    d(s_c s_b s_a ) = -I +O^2\,,
\end{equation*}
and so $\alpha = 1 + O^4$. This means that the second order Taylor term of the
function $\alpha$ at the point $a$ is just zero: $\alpha_2 = 0$.

The Jacobians $j_a, j_b, j_c$ which compose the function $\beta$ in
(\ref{VI.14}) are found in the beginning of Sect.~3. Here one has
\begin{equation*}
    j_{\,q}(\underline{\exp}\,_{q}(\underline{v}\,)) = 1 - \frac 13\, \langle
    \underline{\RR}\,
    \underline{v}\,, \underline{v} \rangle + O^4\,,
\end{equation*}
where $\underline{\RR}$ is defined by (\ref{VI.11}). Therefore
\begin{equation*}
    \beta = 1 - \frac 12\, \langle \underline{\RR}\,(\underline{v}'' - \underline{v}'),
    \underline{v}''-\underline{v}' \rangle + O^4\,.
\end{equation*}
Combining this with (\ref{VI.12}) we conclude that the second
formula in (\ref{VI.10}) holds. $\square$

\begin{lem} The fourth degree contribution $S_4$ in expansion
{\rm(\ref{VI.9})}
satisfies the following estimate
\begin{equation}\label{VI.16}
\langle \pr',\Psi \pr'' \rangle S_4 (v',v'') = O^2(\underline{v}') +
O^2(\underline{v}'')\,.
\end{equation}
\end{lem}

\noindent {\it Proof.} The exponential map corresponding to Christoffel symbols
(\ref{VI.5a}) has the Taylor expansion
\begin{gather*}
    \exp_{a,\xi}(v) = \begin{pmatrix} a + \underline{v} - \frac12\,
    {\underline{\Gamma}}(a)\, {\underline{v}}\,{\underline{v}} + \frac 16\, B(a)\, \underline{v}\,\underline{v}\,\underline{v}
     + O(\underline{v}^4)   \\ \,
     \xi + \underline{\underline{v}} +
     \underline{\Gamma}(a)\,\,\underline{v}\,\underline{\underline{v}} -
     \frac 12\,\big(\xi\,B(a) + C(a)\big)\,\underline{v}\,\underline{v}\, +
     O({\underline{v}}^3) + O({\underline{v}}^2
     \underline{\underline{v}})\,\phantom{\bigg)}
     \end{pmatrix}
\end{gather*}
where the matrices $B,\, C$ are defined in (\ref{VI.5b}), and by
$\underline{\underline{v}}$ we indicate the $p$-component of the vector $v$.

From this formula and (\ref{V.3}) we derive the following expression
for the fourth degree component  $S_4$ of the phase function:
\begin{multline}\label{VI.17}
\ S_4(v',v'') = 2\, \langle \underline{\Gamma}\ \underline{v}'\,
\underline{\underline{v}}'' \,, \,\underline{\Gamma}\ \underline{v}'\,
\underline{v}'' \rangle  - 2\, \langle \underline{\Gamma}\
\underline{v}''\,\underline{\underline{v}}''\,, \,
\underline{\Gamma}\ \underline{v}'\,\underline{v}''  \rangle \,  \phantom{\Big)}\\
+ O\big(\underline{\underline{v}}'\,{\underline{v}''}^{\,3} \big) +
O\big(\underline{\underline{v}}''\,{\underline{v}''}^{\,3} \big) +
O\big(\underline{\underline{v}}'\,{\underline{v}'}^{\,2} \,\underline{v}'' \big)
+ O\big( \underline{\underline{v}}''\,{\underline{v}''}^{\,2} \, \underline{v}'
\big) + O^3 \big(\underline{v}'\,,\underline{v}'' \big)\,.
\end{multline}
Here the remainders do not contribute to the second order derivatives in
$\langle \pr',\Psi \pr'' \rangle\, S_4$ and so we do not need to know their
explicit expression. The first two terms in (\ref{VI.17}) provide formula
(\ref{VI.16}). $\square$

Now we make the rescaling $v={\sqrt \h}\,u$ in the integrand of (\ref{VI.8})
and use Lemma \ref{lem10a}  to get
\begin{multline}\label{VI.16a}
    f \star g \sim \int\!\!\int e^{iS_2(u',u'')} \left [1 + \h\,(iS_4 +
L_2) + O(\h^2) \right ](u',u'')\,\\ \times\,f\big(
\exp_x(\sqrt{\h}u')\big)\, g\big( \exp_x(\sqrt{\h}u'')\big)\,
\frac{du'\, du''}{\pi^{2n}} \,. \qquad \qquad
\end{multline}
By expanding the exponential mappings in (\ref{VI.16a}) we reduce the
calculation of the asymptotics of $f\star g$ to the evaluation of simple
integrals like
\begin{multline}\label{VI.16b}
    \frac 1{\pi^{2n}} \int\!\! \int \P_{2m}(u',u'')\, \exp\big\{2i\langle
    \Omega\,u',u'' \rangle \big\}\, du'\,du'' \\ = \sum_{r=0}^m
    \frac{(-i)^r}{2^r\,  r!} \sum_{\, 1\,\leq\, l_j\, ,\  s_j\, \leq\, 2n }
    \Psi^{l_1\, s_1} \cdots \Psi^{l_r\, s_r }\big( \pr_{l_1}'\dots
    \pr'_{l_r}\, \pr''_{s_1} \dots \pr''_{s_r}\P_{2m} \big)(0,0)
\end{multline}
with some polynomials, $\P_{2m}$, of degree $2m$, where $m=0,1,\dots$.

In order to know the $k^{th}$ term in expansion (\ref{VI.1}) one must take into
account contributions of integrals like (\ref{VI.16b}) for degrees $2m\leq 2k$.
In this way from (\ref{VI.16a}) we obtain the expansion (\ref{VI.3}) and the
formula (\ref{VI.4}) for the operator $G_2$, where the tensor $\B$ is given by
\begin{equation}\label{VI.18}
    \B^{sl} = 2\, \Psi^{sm}\, \Psi^{rl}\, \pr_m'\,\pr_r'' L_2 + \Psi^{sm}\, \Psi^{rl}\,
    \Psi^{jk}\, \pr_m'\,\pr_j'\,\pr_r''\,\pr_k''\,S_4\,.
\end{equation}
The second formula in (\ref{VI.10}) implies that
\begin{equation*}
    \pr'\,\pr'' L_2 = \begin{pmatrix} \underline{\RR} & 0 \\ 0
    & 0 \end{pmatrix}\,,
\end{equation*}
where $\underline{\RR}$ is given by (\ref{VI.11}). From (\ref{VI.16}) one
concludes that the second term in (\ref{VI.18}) vanishes. Therefore
\begin{equation}\label{VI.19}
    \B = 2\, \Psi\cdot \begin{pmatrix} \underline{\RR} & 0 \\ 0
    & 0 \end{pmatrix}\cdot \Psi\,.
\end{equation}

\begin{prop}\label{prop5} The curvature tensor $R$ of the magneto-geodesic
connection $\nabla$ on $\TB$ is given by
\begin{align*} \label{VI.20}
R_{q^i q^j q^k}^{q^s} &= \underline{R}_{\,ijk}^{\,s} \,,
\\
R_{p_j q^i q^k}^{p_s} &= \underline{R}_{\,ski}^{\,j}\,, \qquad \quad
R_{q^i p_j q^k}^{p_s} = -R_{q^i q^k p_j }^{p_s}
= \frac 13 \, \big(\underline{R}^j_{\,ski}
 + \underline{R}^j_{\,iks}\big)\,, \phantom{\Bigg)}
\\
R_{q^i q^j q^k}^{p_s} &=\frac 13 \,
 \,  \underset{i\,s}{\mathfrak S}\,  p_m \Big
(\underline{\nabla}_{\,s}\, \underline{R}_{\ ikj}^{\,m} + 3\,
\underline{\Gamma}_{\ sl}^{\,m}\,  \underline{R}^{\,l}_{\ ijk} -
\underline{\Gamma}_{\ jl}^{\,m}\,  \underline{R}^{\,l}_{\ iks} +
\underline{\Gamma}_{\ kl}^{\,m}\,  \underline{R}^{\,l}_{\ ijs} \Big) +
M_{sijk}\,,
\end{align*}
where
\begin{equation*}
    M_{sijk} \df\  \frac13\, \Big(\underline{\nabla}_{\,i}
\underline{\nabla}_{\,s}\,F_{\,jk} + 2\, \underline{R}^{\,l}_{\ ijk} \,F_{\,ls}
+ \underset{ijk}{\mathfrak S}\,\underline{R}^{\,l}_{\ sij} \,F_{\,lk} \Big)\,,
\quad {(magnetic\ curvature)}\,.
\end{equation*}
Here $\underline{R} = \underline{R}(q)$ and $F = F(q)$ are  the curvature of
the connection $\underline{\nabla}$ and the magnetic tensor on $\M$. All other
components of the curvature $R$ vanish.
\end{prop}

We first remark that the block $R^p_{qqq}$ of the curvature tensor $R$ is not
itself a tensor. But the part $M$, which we call the \emph{magnetic curvature},
is a tensor on $\M$. This part of the total curvature entangles the magnetic
field $F$ with the curvature tensor $\underline{R}$ on $\M$. Also note that the
only $p$-dependent part of the curvature is the $R^p_{qqq}$ block (where $p$
enters linearly); all other parts are strictly $q$-dependent.

For any symplectic connection,
the Ricci tensor
\begin{equation}\label{VI.18a}
    \RR_{\,ij} \df R_{\,ikj}^{\,k}
\end{equation}
is symmetric \cite{Vais85}.
Thus for the magneto-geodesic connection on $T^*\M$
we have $\RR_{\,ij}=\RR_{\,ji}$.

\begin{cor}
The Ricci tensor
for the magneto-geodesic connection on $T^*\M$
is given by
\begin{equation}\label{VI.18b}
    \RR = \frac 23\begin{pmatrix} \underline{\RR} \phantom{\Big )}& 0 \\
    0  & 0 \end{pmatrix}\,,
\end{equation}
where $\underline{\RR}$ is the symmetric part
of the Ricci tensor for the affine connection on $\M$.
\end{cor}

Now one just has to compare this formula for the Ricci tensor
with formula (\ref{VI.19}) for the tensor $\B$ in the
$\h$-expansion of the $\star$-product (\ref{VI.3}), (\ref{VI.4}).

Altogether we have proved the following statement.

\begin{theor} \label{thm2} The magneto-geodesic product {\rm{(\ref{I.5})}} over $\TB$ has the
asymptotic expansion {\rm(\ref{VI.3})} with the second order term $G_2$ given
by the formula {\rm(\ref{VI.4})}. In this formula the connection $\nabla$
coincides with the magneto-geodesic connection {\rm(\ref{VI.5}), (\ref{VI.5a})}
and the tensor $\B$ is
\begin{equation}
    \B = 3\,\Psi \RR \Psi\,,
\end{equation} where $\RR$ is the Ricci tensor of this
connection.
\end{theor}

The magneto-geodesic connection $\nabla$ on $\TB$ entangles, on the phase space
level, the original affine connection $\underline{\nabla}$ and the magnetic
field (Faraday tensor) $F$ on $\M$.  In the case of zero magnetic field $F=0$
this connection was obtained in \cite{BNW98} via a version of
the deformation quantization approach by using
a different star-product which does not obey axioms (\ref{II.6}).

The part of the Christoffel symbol in (\ref{VI.5a})
which describes the `interaction' between the affine and magnetic structures on
$\M$ is given by the tensor $C$. The tensor $C$ contains the symmetrized
covariant derivative of the magnetic tensor $F$. We label this part of the
connection as the \emph{magneto-geodesic coupling}} tensor. The usual covariant
derivative tensor $\underline{\nabla} F$ we call the \emph{magnetic
inhomogeneity}.

Let us examine the symmetric and skew-symmetric properties of these two
tensors. We represent this pair of tensors with the notation
\begin{equation*}
    F_{[jk]l} \df \underline{\nabla}_{\,l}\, F_{jk}\,, \qquad  F_{j\{kl\} } \df C_{jkl}\,.
\end{equation*}

\begin{prop}
{\rm(a)} The magnetic inhomogeneity  tensor $F_{[jk]l}$ is skew-symmetric in
the first pair of indices; the magneto-geodesic coupling tensor $F_{j\{kl\}}$
is symmetric in the last pair of indices. They are related to each other by the
following duality formulas
\begin{equation*}
F_{j\{kl\}} = {\frac 13} \big( F_{[jk]l} + F_{[jl]k} \big)\,,
\qquad F_{[jk]l} = F_{j\{kl\}} - F_{k\{jl\}} \,.
\end{equation*}

\noindent {\rm(b)} Both of these tensors obey the cyclic property
\begin{equation}\label{VI.20}
    \underset{jkl}{\mathfrak S}\, F_{j\{kl\}} = 0\,,  \qquad
    \underset{jkl}{\mathfrak S}\,  F_{[jk]l} = 0\,.
\end{equation}

\noindent{\rm(c)} The magneto-geodesic coupling tensor is related to the
magnetic curvature tensor $M$ by
\begin{equation*}
    \underline{\nabla}_j F_{s\{ik\} } - \underline{\nabla}_k F_{s\{ij\} } =
    M_{isjk}\,.
\end{equation*}

\noindent{\rm(d)} On Riemannian manifolds $\M$ {\rm(}with the Levi-Civita connection
$\underline{\nabla}$ defined by metric $g${\rm)} the current
covector
\begin{equation*}
    j_s = \frac 23 F_{s\{ik\} }\, g^{ik}
\end{equation*} satisfies the continuity equation
$g^{ls}\underline{\nabla}_{\ l} j_s = 0$, and generates the exact
{\rm 2}-form $\varkappa \equiv dj$ with coefficients
\begin{equation*}
    \varkappa_{sm} = g^{ik}\big( 3 M_{iksm} + 2 F_{il}\, \underline{R}_{ksm}^l \big)
    \,.
\end{equation*} The identity $d\varkappa = 0$, or
${\mathfrak S}\, \nabla_r \varkappa_{sm} = 0$, ties together the
magnetic curvature $M$ and the Riemannian curvature $\underline{R}$ on $\M$.
\end{prop}

The last statement in (\ref{VI.20}) is just the homogeneous Maxwell equation
for the magnetic field (that ensures that there are no Dirac monopoles), and
the first identity in (\ref{VI.20}) is the `dual' to the Maxwell equation and
incorporates the magneto-geodesic coupling tensor.

Note that within the class of tensors obeying the condition ${\mathfrak S}\,
C_{jkl} = 0$, formulas (\ref{VI.5a}), (\ref{VI.5b}) determine a unique
connection on $\TB$ having the property that the symplectic form is covariantly
constant, $\nabla \omega = 0$.

In the Euclidean case $\M=\R^n$ the magneto-geodesic connection $\nabla$
coincides with the magnetic connection found in \cite{KO3}. In this case the
curvature tensor $R$ of $\nabla$ (see Proposition 5) is reduced to the magnetic
curvature which is just the second derivative matrix $M_{sijk} =  \frac 13\,
D^2_{si} F_{jk}$. Thus the curvature $R$ is constant iff the field $F$ is
quadratic in Euclidean coordinates. In this way we obtained in \cite{KO3} an
example of a symmetric symplectic space which can be explicitly and exactly
quantized.

It should be interesting to study the equation
\begin{equation}\label{VI.21}
    \nabla R = 0
\end{equation}
for the general magnetic-geodesic connection. Under condition (\ref{VI.21}) one
again has an example of a symmetric symplectic structure, now on $\TB$, which
is explicitly and exactly (not formally and not asymptotically)
quantized.
It would be interesting to compare this construction of
quantized symmetric spaces with \cite{BCG95,Biel02}.

\bigskip {\bf Acknowledgments}. The first author is grateful to Russian Foundation
for Basic Research for partial support (grant No.~05-01-00918-a). The
research of T.A.O. is supported by a grant from Natural Sciences and
Engineering Research Council of Canada. The authors thank the
Winnipeg Institute of Theoretical Physics for its continuing support.

\renewcommand{\theequation}{\Alph{section}.\arabic{equation}}

\end{document}